%% file: provablepaper.tex
\documentclass[sigplan,10pt]{acmart}
\renewcommand\footnotetextcopyrightpermission[1]{}
\usepackage{amsmath}
\usepackage{listings}
\usepackage{cuted}
\usepackage{capt-of}
\input{listings-scheme}

\setcopyright{acmlicensed}
\copyrightyear{2025}
\acmYear{2025}
\acmDOI{XXXXXXX.XXXXXXX}

\begin{document}

\title{Shock with Confidence: Formal Proofs of Correctness for Hyperbolic Partial Differential Equation Solvers}

\author{Jonathan Gorard}
\affiliation{\institution{Princeton University}
\city{Princeton, NJ}
\country{USA}}
\email{gorard@princeton.edu}

\author{Ammar Hakim}
\affiliation{\institution{Princeton Plasma Physics Laboratory}
\city{Princeton, NJ}
\country{USA}}
\email{ahakim@pppl.gov}

\renewcommand{\shortauthors}{Gorard and Hakim}
\renewcommand{\shorttitle}{Shock with Confidence}

\begin{abstract}
First-order systems of hyperbolic partial differential equations (PDEs) occur ubiquitously throughout computational physics, commonly used in simulations of fluid turbulence, shock waves, electromagnetic interactions, and even general relativistic phenomena. Such equations are often challenging to solve numerically in the non-linear case, due to their tendency to form discontinuities even for smooth initial data, which can cause numerical algorithms to become unstable, violate conservation laws, or converge to physically incorrect solutions. In this paper, we introduce a new formal verification pipeline for such algorithms in Racket, which allows a user to construct a bespoke hyperbolic PDE solver for a specified equation system, generate low-level C code which verifiably implements that solver, and then produce formal proofs of various mathematical and physical correctness properties of the resulting implementation, including ${L^2}$ stability, flux conservation, and physical validity. We outline how these correctness proofs are generated, using a custom-built theorem-proving and automatic differentiation framework that fully respects the algebraic structure of floating-point arithmetic, and show how the resulting C code may either be used to run standalone simulations, or integrated into a larger computational multiphysics framework such as \textsc{Gkeyll}.
\end{abstract}

\maketitle

\lstset{language=Scheme}

\section{Introduction}

Systems of hyperbolic partial differential equations (PDEs) constitute a highly general class of mathematical models for phenomena involving waves, or the wave-like propagation of information, including water waves in hydrodynamic modeling\cite{vreugdenhil_numerical_1994}, blast waves in explosives modeling\cite{fickett_introduction_1985}, gravitational waves in general relativity\cite{auger_overview_2017}\footnote{Indeed, the Einstein field equations themselves can be cast in a purely hyperbolic form\cite{gorard_gravitas_2024}.}, and even traffic waves in vehicular traffic flows\cite{musha_traffic_1978}. When systems of hyperbolic PDEs are non-linear, they have a tendency to form discontinuities after finite time (even for arbitrarily smooth initial data)\cite{toro_centred_2000}, such as \textit{shock waves} and \textit{contact discontinuities}. Such discontinuities are notoriously challenging to capture using standard numerical techniques such as finite difference methods, since the discontinuous solution to the PDE system does not exist in the strong sense (only in the \textit{weak}/\textit{distributional} sense), and such methods will therefore typically violate certain crucial conservation laws (such as conservation of energy, or conservation of momentum) in the presence of such discontinuities\cite{harten_finitedifference_1976}. This has led to the development of a variety of \textit{finite volume} algorithms, and specifically \textit{high-resolution shock-capturing} (HRSC) schemes\cite{toro_riemann_2009}, which capture these non-linear discontinuities by solving the \textit{weak}/\textit{integral} form of the PDE system directly, in a manner which fully guarantees conservation. However, Godunov's theorem\footnote{Linear, monotonicity-preserving schemes cannot be more than first-order accurate\cite{laney_computational_1998}.}, a central result in the theory of numerical methods for hyperbolic PDEs, implies that naive algorithms have a tendency to produce spurious oscillations (i.e. violations of the \textit{total variation diminishing}, or TVD, property\cite{harten_high_1983}), or otherwise to introduce new extrema into the solution (i.e. violations of the \textit{monotonicity-preservation} property). One way around this highly distressing theorem is to sacrifice either linearity or higher-order accuracy of the method in the vicinity of shock waves. To make matters worse, since thermodynamic quantities such as entropy tend to be discontinuous across shocks, some methods may converge to mathematically correct but physically invalid solutions (e.g. solutions which decrease total entropy), unless certain additional \textit{entropy conditions} are satisfied\cite{osher_riemann_1984}.

In this paper, we introduce a new technique for producing \textit{formally verified implementations} of high-resolution shock-capturing schemes, with provable mathematical and physical correctness properties. We develop a domain-specific language in Racket\cite{felleisen_et_al} for representing arbitrary first-order systems of hyperbolic PDEs, from which verifiable C code can then be automatically synthesized, implementing a variety of bespoke first- and second-order accurate finite volume algorithms to solve them. We also introduce a suite of novel automated theorem-proving and automatic differentiation tools which fully respect the algebraic properties of floating-point arithmetic (e.g. allowing commutativity but not associativity of floating-point addition and multiplication), and which thus implement only \textit{strictly correctness-preserving} algebraic transformations of the corresponding C code. We show how it is possible to use these tools to prove fundamental correctness properties of the underlying first-order numerical schemes, such as (strict) hyperbolicity-preservation, ${L^1}$, ${L^2}$ and ${L^{\infty}}$ stability, local Lipschitz continuity (a sufficient condition for physical correctness), and flux conservation. We also demonstrate how it is possible to prove correctness properties of the flux extrapolation algorithms necessary to extend these schemes to second-order spatial accuracy, such as symmetry and TVD. Each generated proof is, itself, a symbolic piece of Racket code that can be independently run and verified, and is therefore able to act as a standalone \textit{certificate of correctness} for some aspect of the algorithm. This pipeline is also able to produce both entirely standalone verified solvers, as well as verified solvers that can be integrated into a larger computational multiphysics framework such as \textsc{Gkeyll}\footnote{\url{https://gkeyll.readthedocs.io/en/latest/}} (with the larger framework handling all non-verified simulation tasks, such as parallelism, grid generation, and data input/output). The underlying theory behind the finite volume solvers used within \textsc{Gkeyll} can be found in \cite{hakim_high_2006}.

Section \ref{sec:sec1} begins with preliminary background material on the mathematical properties of high-resolution shock-capturing schemes. In \ref{sec:sec2}, we briefly introduce explicitly conservative finite volume numerical methods, as well as the four model equation systems (linear advection, inviscid Burgers', perfectly hyperbolic Maxwell's, and isothermal Euler) that will be used throughout the paper. In \ref{sec:sec3}, we introduce the Lax-Friedrichs inter-cell flux function, and outline sufficient conditions to guarantee its hyperbolicity-preservation, CFL stability, and local Lipschitz continuity (with the latter condition itself being a sufficient condition for physical/thermodynamic correctness). In \ref{sec:sec4}, we introduce the Roe approximate Riemann solver, and again outline sufficient conditions to guarantee its hyperbolicity-preservation and flux conservation (i.e. jump continuity). In \ref{sec:sec5}, we describe how to extend the first-order Lax-Friedrichs and Roe solvers to be second-order accurate in space via flux extrapolation, outline sufficient conditions on the flux limiters to guarantee that the resulting second-order schemes be symmetric and total variation diminishing (TVD), and also introduce the four flux limiters (minmod, monotonized-centered, superbee and van Leer) that are used throughout the paper. 

Section \ref{sec:sec6} presents the main methodology and results of the paper. In \ref{sec:sec7}, we outline how we encode hyperbolic PDE systems in Racket, and how the automatic C code generator works. In \ref{sec:sec8}, we introduce the core of the symbolic theorem-prover, outlining the algorithm for reducing an arbitrary symbolic Racket expression to a normal form, whilst respecting the algebraic structure of floating-point arithmetic. In \ref{sec:sec9}, we describe the algorithm for performing automatic differentiation on arbitrary Racket expressions, enabling one to compute symbolic gradients, Jacobians, Hessians, etc. In \ref{sec:sec10}, we outline how these methods may be combined to produce automatic proofs of (strict) hyperbolicity, CFL stability, convexity/local Lipschitz continuity, and flux continuity properties of Lax-Friedrichs and Roe solvers. In \ref{sec:sec11}, we describe the algorithm for performing variable transformations and evaluating symbolic limits of Racket expressions for the purpose of proving symmetry and second-order TVD properties of flux limiters. 

In \ref{sec:sec12}, we present our main results: full proofs of correctness for Lax-Friedrichs and Roe solvers for the linear advection, inviscid Burgers', and perfectly hyperbolic Maxwell's equations, with partial/conditional proofs of correctness in the case of the isothermal Euler equations, along with full proofs of correctness for the minmod and monotonized-centered flux limiters, with partial proofs of correctness for the superbee and van Leer limiters. We end in Section \ref{sec:sec13} with some concluding remarks and directions for future research.

Our Racket implementation of the formal verification pipeline can be found at\footnote{\url{https://github.com/ammarhakim/gkylcas/}}. The corresponding provably-correct C implementations, which have been integrated into the \textsc{Gkeyll} code, can be found at\footnote{\url{https://github.com/ammarhakim/gkylzero/}}.

\section{Preliminaries}
\label{sec:sec1}

\subsection{Hyperbolic PDE Solvers}
\label{sec:sec2}

For the purposes of this paper, we consider homogeneous systems of (potentially non-linear) first-order hyperbolic PDEs, expressed in the generic \textit{conservation law} form:

\begin{equation}
\frac{\partial \mathbf{U}}{\partial t} + \nabla \cdot \mathbf{F} \left( \mathbf{U} \right) = \mathbf{0},
\end{equation}
where ${\mathbf{U}}$ represents the vector of \textit{conserved variables} and ${\mathbf{F} \left( \mathbf{U} \right)}$ represents the matrix of \textit{fluxes} of those variables. We can discretize such a system of PDEs in both space and time, using a uniform grid spacing of ${\Delta x}$ and a time-step of ${\Delta t}$, indexed by subscripts ($i$) and superscripts ($n$):

\begin{equation}
x_{i + 1} = x_i + \Delta x, \qquad \text{ and } \qquad t^{n + 1} = t^n + \Delta t,
\end{equation}
respectively. Specifically, we employ a \textit{finite volume} discretization, in which the conserved variable vector ${\mathbf{U}}$ at time ${t^n}$ is taken to be constant within each cell ${x_i}$ (and therefore piecewise constant across the entire domain), and we adopt the shorthand ${\mathbf{U}_{i}^{n} = \mathbf{U} \left( x_i, t^n \right)}$ to denote this value. The value of the conserved variable vector ${\mathbf{U}_{i}^{n}}$ within each cell can then be evolved in time by means of the explicitly conservative update formula\cite{toro_riemann_2009}:

\begin{equation}
\mathbf{U}_{i}^{n + 1} = \mathbf{U}_{i}^{n} - \frac{\Delta t}{\Delta x} \left[ \mathbf{F}_{i + \frac{1}{2}} - \mathbf{F}_{i - \frac{1}{2}} \right],
\end{equation}
where ${\mathbf{F}_{i + \frac{1}{2}}}$ and ${\mathbf{F}_{i - \frac{1}{2}}}$ represent \textit{inter-cell fluxes}, i.e. the interpolated fluxes of the conserved variables between cells ${x_i}$ and ${x_{i + 1}}$, and between cells ${x_{i - 1}}$ and ${x_i}$, respectively. Different choices of inter-cell flux function give rise to different finite volume schemes, exhibiting different numerical properties\cite{leveque_finite_2011}.

In what follows, we shall focus on four one-dimensional model equation systems in particular, covering all possible combinations of linear vs. non-linear and scalar vs. vector. The first two are purely 1D scalar equations of the form:

\begin{equation}
\frac{\partial u}{\partial t} + \frac{\partial f \left( u \right)}{\partial x} = 0,
\end{equation}
namely the linear \textit{advection equation} with flux ${f \left( u \right) = a u}$ (for an arbitrary constant advection velocity ${a \in \mathbb{R}}$), and the non-linear \textit{inviscid Burgers' equation} with flux ${f \left( u \right) = \frac{1}{2} u^2}$. In each case, the conserved variable $u$ simply represents an arbitrary advected quantity. The second two are 1D vector equation systems of the form:

\begin{equation}
\frac{\partial \mathbf{U}}{\partial t} + \frac{\partial \mathbf{F} \left( \mathbf{U} \right)}{\partial x} = \mathbf{0}
\end{equation}
namely the linear \textit{perfectly hyperbolic Maxwell's equations}\cite{munz_divergence_2000}, with conserved variable vector:

\begin{equation}
\begingroup
\setlength\arraycolsep{3pt}
\mathbf{U} = \begin{bmatrix}
E^x & E^y & E^z & B^x & B^y & B^z & \phi & \psi
\end{bmatrix}^{\intercal},
\endgroup
\end{equation}
and flux vector:

\begin{equation}
\begingroup
\setlength\arraycolsep{3pt}
\mathbf{F} \left( \mathbf{U} \right) = \begin{bmatrix}
\chi c^2 \phi & c^2 B^z & - c^2 B^y & \gamma \psi & - E^z & E^z & \chi E^x & \gamma c^2 B^x
\end{bmatrix}^{\intercal},
\endgroup
\end{equation}
and the non-linear \textit{isothermal Euler equations}, with conserved variable vector:

\begin{equation}
\begingroup
\setlength\arraycolsep{3pt}
\mathbf{U} = \begin{bmatrix}
\rho & \rho u & \rho v & \rho w
\end{bmatrix}^{\intercal},
\endgroup
\end{equation}
and flux vector:

\begin{equation}
\begingroup
\setlength\arraycolsep{3pt}
\mathbf{F} \left( \mathbf{U} \right) = \begin{bmatrix}
\rho u & \rho u^2 + \rho v_{th}^{2} & \rho u v & \rho u w
\end{bmatrix}^{\intercal}.
\endgroup
\end{equation}
For the perfectly hyperbolic Maxwell's equations, the conserved variables correspond to the $x$, $y$ and $z$ components of the electric and magnetic field vectors ${\mathbf{E}}$ and ${\mathbf{B}}$, as well as the correction potentials ${\phi}$ and ${\psi}$ for cleaning divergence errors in ${\mathbf{E}}$ and ${\mathbf{B}}$, respectively. ${c, \chi, \gamma \in \mathbb{R}}$ are arbitrary constants representing the speed of light, and the propagation speeds of electric and magnetic divergence errors, respectively. For the isothermal Euler equations, the conserved variables correspond to the fluid density ${\rho}$, and the $x$, $y$ and $z$ components of the fluid momentum ${\rho u}$, ${\rho v}$ and ${\rho w}$ (with $u$, $v$ and $w$ representing the $x$, $y$ and $z$ components of the fluid velocity, which are not conserved). ${v_{th} \in \mathbb{R}}$ is an arbitrary constant representing the thermal velocity of the fluid.

\subsection{The Lax-Friedrichs Flux}
\label{sec:sec3}

One of the simplest choices of inter-cell flux function is the Lax-Friedrichs (finite difference) flux\cite{lax_weak_1954}\cite{leveque_numerical_1992}:

\begin{equation}
\mathbf{F}_{i + \frac{1}{2}} = \frac{1}{2} \left[ \mathbf{F} \left( \mathbf{U}_{i}^{n} \right) + \mathbf{F} \left( \mathbf{U}_{i + 1}^{n} \right) \right] - \frac{\Delta x}{2 \Delta t} \left[ \mathbf{U}_{i + 1}^{n} - \mathbf{U}_{i}^{n} \right].
\end{equation}
A finite volume solver based on a Lax-Friedrichs inter-cell flux will satisfy the following properties\cite{breus_correct_2004}:

\begin{theorem}
\label{th:lax1}
A Lax-Friedrichs solver preserves hyperbolicity\cite{hartman_lemma_1960} (i.e. preserves well-posedness of the Cauchy problem along all non-characteristic hypersurfaces) for an arbitrary hyperbolic PDE system if the flux Jacobian with respect to the conserved variable vector ${\mathbf{U}}$:

\begin{equation}
\mathbf{J}_{\mathbf{F}} = \nabla_{\mathbf{U}} \mathbf{F} \left( \mathbf{U} \right),
\end{equation}
is diagonalizable, with purely real eigenvalues.
\end{theorem}

\begin{theorem}
\label{th:lax2}
A Lax-Friedrichs solver maintains ${L^1}$, ${L^2}$ and ${L^{\infty}}$ stability for an arbitrary hyperbolic PDE system if the CFL stability condition\cite{courant_partial_1967}:

\begin{equation}
0 \leq \frac{\left\lvert a \right\rvert \Delta t}{\Delta x} \leq 1,
\end{equation}
is satisfied, where ${\left\lvert a \right\rvert}$ denotes the largest absolute eigenvalue of the flux Jacobian ${\mathbf{J}_{\mathbf{F}}}$.
\end{theorem}

\begin{theorem}
\label{th:lax3}
A Lax-Friedrichs solver preserves local Lipschitz continuity of the discrete flux function ${\mathbf{F} \left( \mathbf{U}_{i}^{n} \right)}$ with respect to the discrete conserved variables ${\mathbf{U}_{i}^{n}}$ (i.e. preserves the property that, in every neighborhood ${\mathcal{N}}$ of possible values of ${\mathbf{U}_{i}^{n}}$, there exists a restriction of the discrete flux function to ${\mathcal{N}}$ which has strictly bounded first derivatives in ${\mathbf{U}_{i}^{n}}$) for an arbitrary hyperbolic PDE system if the continuous flux function ${\mathbf{F} \left( \mathbf{U} \right)}$ is itself locally Lipschitz continuous with respect to the continuous conserved variables ${\mathbf{U}}$.
\end{theorem}

\begin{theorem}
\label{th:lax4}
A sufficient condition for the continuous flux function ${\mathbf{F} \left( \mathbf{U} \right)}$ for an arbitrary hyperbolic PDE system to be locally Lipschitz continuous with respect to the continuous conserved variables ${\mathbf{U}}$ is for the componentwise Hessian with respect to ${\mathbf{U}}$:

\begin{equation}
\mathbf{H}_f = \left( \nabla_{\mathbf{U}} \nabla_{\mathbf{U}} f \left( \mathbf{U} \right) \right)^{\intercal}
\end{equation}
to be positive semidefinite (i.e. symmetric/Hermitian with non-negative eigenvalues), for each scalar flux component ${f \left( \mathbf{U} \right)}$ of ${\mathbf{F} \left( \mathbf{U} \right)}$.
\end{theorem}

The significance of hyperbolicity-preservation is that it is a necessary condition for the solver to remain deterministic, i.e. to prevent discrete solutions from becoming multi-valued\cite{hartman_lemma_1960}. Note that the flux Jacobian ${\mathbf{J}_{\mathbf{F}}}$ being symmetric/Hermitian is a sufficient but not necessary condition for the solver to preserve hyperbolicity. The significance of the CFL stability condition is that, at least for a Lax-Friedrichs solver, CFL stability subsumes all other standard notions of numerical stability (e.g. ${L^1}$, ${L^2}$ and ${L^{\infty}}$ stability) within a single condition\cite{courant_partial_1967}. Finally, the significance of the local Lipschitz continuity condition, as well as the slightly stronger convexity condition, is that these constraints are sufficient to guarantee that the solver will converge to the physically correct (i.e. thermodynamically consistent) solution in the presence of weak, nonlinear waves, such as shock waves in hydrodynamics\cite{breus_correct_2004}.

\subsection{The Roe Flux}
\label{sec:sec4}

A more sophisticated choice of inter-cell flux function is the Roe (linearized Riemann problem) flux\cite{roe_approximate_1981}:

\begin{equation}
\mathbf{F}_{i + \frac{1}{2}} = \frac{1}{2} \left[ \mathbf{F} \left( \mathbf{U}_{i}^{n} \right) + \mathbf{F} \left( \mathbf{U}_{i + 1}^{n} \right) \right] - \frac{1}{2} \sum_p \left\lvert \lambda_p \right\rvert \alpha_p \mathbf{r}_p.
\end{equation}
In the above, we assume an inter-cell \textit{Roe matrix} ${\mathbf{A} \left( \mathbf{U}_{i}^{n}, \mathbf{U}_{i + 1}^{n} \right)}$, which is a linearized approximation to the flux Jacobian ${\mathbf{J}_{\mathbf{F}}}$ that is taken to be constant between cells ${x_i}$ and ${x_{i + 1}}$, satisfying consistency with the exact Jacobian in the appropriate limit:

\begin{equation}
\lim_{\mathbf{U}_{i}^{n}, \mathbf{U}_{i + 1}^{n} \to \mathbf{U}} \left[ \mathbf{A} \left( \mathbf{U}_{i}^{n}, \mathbf{U}_{i + 1}^{n} \right) \right] = \nabla_{\mathbf{U}} \mathbf{F} \left( \mathbf{U} \right),
\end{equation}
such that ${\lambda_p}$ are the eigenvalues of ${\mathbf{A} \left( \mathbf{U}_{i}^{n}, \mathbf{U}_{i + 1}^{n} \right)}$, ${\mathbf{r}_p}$ are the (right) eigenvectors of ${\mathbf{A} \left( \mathbf{U}_{i}^{n}, \mathbf{U}_{i + 1}^{n} \right)}$, and ${\alpha_p}$ are the components of the inter-cell jump in the conserved variables ${\mathbf{U}_{i + 1}^{n} - \mathbf{U}_{i}^{n}}$, as represented in the ${\mathbf{r}_p}$ eigenbasis:

\begin{equation}
\mathbf{U}_{i + 1}^{n} - \mathbf{U}_{i}^{n} = \sum_p \alpha_p \mathbf{r}_p.
\end{equation}
A finite volume solver based on a Roe inter-cell flux will satisfy the following properties\cite{wesseling_principles_2001}:

\begin{theorem}
\label{th:roe1}
A Roe solver preserves hyperbolicity for an arbitrary hyperbolic PDE system if the Roe matrix ${\mathbf{A} \left( \mathbf{U}_{i}^{n}, \mathbf{U}_{i + 1}^{n} \right)}$ is diagonalizable, with purely real eigenvalues.
\end{theorem}

\begin{theorem}
\label{th:roe2}
A Roe solver is flux-conservative (i.e. exactly conserves the components of the conserved variable vector ${\mathbf{U}}$) if it satisfies the flux jump condition:

\begin{equation}
\mathbf{F} \left( \mathbf{U}_{i + 1}^{n} - \mathbf{U}_{i}^{n} \right) = \mathbf{A} \left( \mathbf{U}_{i}^{n}, \mathbf{U}_{i + 1}^{n} \right) \left[ \mathbf{U}_{i + 1}^{n} - \mathbf{U}_{i}^{n} \right].
\end{equation}
\end{theorem}

Note that Theorem \ref{th:roe1} regarding Roe solvers is essentially identical in content to Theorem \ref{th:lax1} regarding Lax-Friedrichs solvers, but with the exact flux Jacobian ${\mathbf{J}_{\mathbf{F}}}$ replaced by the linearized inter-cell approximation ${\mathbf{A} \left( \mathbf{U}_{i}^{n}, \mathbf{U}_{i + 1}^{n} \right)}$. Note, moreover, that a Lax-Friedrichs solver is guaranteed to conserve the components of ${\mathbf{U}}$ automatically, and does not require any additional conditions (such as those in Theorem \ref{th:roe2}) to hold. Finally, note that the CFL stability and local Lipschitz continuity criteria for Lax-Friedrichs solvers apply to Roe solvers too, i.e. a Roe solver for a given PDE system will not be CFL stable unless the corresponding Lax-Friedrichs solver is also CFL stable, etc.

\subsection{Flux Extrapolation and Limiters}
\label{sec:sec5}

The Lax-Friedrichs and Roe solvers described above are, naively, only first-order accurate in space. In order to achieve second-order spatial accuracy, it is necessary to replace the piecewise constant approximations of the conserved variable vector ${\mathbf{U}}$ with piecewise linear approximations instead\cite{van_leer_towards_1979}. At the boundary between cells ${x_i}$ and ${x_{i + 1}}$, the left- and right-sided extrapolated values of ${\mathbf{U}}$ are given by:

\begin{equation}
\mathbf{U}_{i + \frac{1}{2}}^{L} = \mathbf{U}_{i} + \frac{1}{2} \boldsymbol\phi \left( \mathbf{r}_i \right) \left( \mathbf{U}_{i + 1} - \mathbf{U}_i \right),
\end{equation}
and:

\begin{equation}
\mathbf{U}_{i + \frac{1}{2}}^{R} = \mathbf{U}_{i + 1} - \frac{1}{2} \boldsymbol\phi \left( \mathbf{r}_{i + 1} \right) \left( \mathbf{U}_{i + 2} - \mathbf{U}_{i + 1} \right),
\end{equation}
respectively, while at the boundary between cells ${x_{i - 1}}$ and ${x_i}$, the left- and right-sided extrapolated values of ${\mathbf{U}}$ are given by:

\begin{equation}
\mathbf{U}_{i - \frac{1}{2}}^{L} = \mathbf{U}_{i - 1} + \frac{1}{2} \boldsymbol\phi \left( \mathbf{r}_{i - 1} \right) \left( \mathbf{U}_i - \mathbf{U}_{i - 1} \right),
\end{equation}
and:

\begin{equation}
\mathbf{U}_{i - \frac{1}{2}}^{R} = \mathbf{U}_i - \frac{1}{2} \boldsymbol\phi \left( \mathbf{r}_i \right) \left( \mathbf{U}_{i + 1} - \mathbf{U}_i \right),
\end{equation}
respectively, where ${\mathbf{r}_i}$ is a ratio of successive gradients of ${\mathbf{U}}$ (computed componentwise):

\begin{equation}
\mathbf{r}_i = \frac{\mathbf{U}_i - \mathbf{U}_{i - 1}}{\mathbf{U}_{i + 1} - \mathbf{U}_i}.
\end{equation}
The function ${\boldsymbol\phi \left( \mathbf{r}_i \right)}$ in the above is a \textit{flux limiter}\cite{leveque_finite_2011}, intended to limit the spatial derivatives arising within the reconstruction so as to damp any spurious oscillations that might otherwise appear in the vicinity of steep gradients or true discontinuities. Since all quantities are computed componentwise, it is sufficient to think of the flux limiter as being a scalar function of a scalar ratio of gradients ${\phi \left( r \right)}$.

A second-order spatially accurate scheme can now be constructed by taking the left- and right-extrapolated values ${\mathbf{U}_{i + \frac{1}{2}}^{L}}$ and ${\mathbf{U}_{i + \frac{1}{2}}^{R}}$ and evolving them forward by a half time-step ${\frac{\Delta t}{2}}$, to obtain the evolved states ${\overline{\mathbf{U}_{i + \frac{1}{2}}^{L}}}$ and ${\overline{\mathbf{U}_{i + \frac{1}{2}}^{R}}}$:

\begin{equation}
\overline{\mathbf{U}_{i + \frac{1}{2}}^{L}} = \mathbf{U}_{i + \frac{1}{2}}^{L} + \frac{\Delta t}{2 \Delta x} \left[ \mathbf{F} \left( \mathbf{U}_{i - \frac{1}{2}}^{R} \right) - \mathbf{F} \left( \mathbf{U}_{i + \frac{1}{2}}^{L} \right) \right],
\end{equation}
and:

\begin{equation}
\overline{\mathbf{U}_{i + \frac{1}{2}}^{R}} = \mathbf{U}_{i + \frac{1}{2}}^{R} + \frac{\Delta t}{2 \Delta x} \left[ \mathbf{F} \left( \mathbf{U}_{i + \frac{1}{2}}^{R} \right) - \mathbf{F} \left( \mathbf{U}_{i + \frac{3}{2}}^{L} \right) \right],
\end{equation}
respectively. The inter-cell flux ${\mathbf{F}_{i + \frac{1}{2}}}$ for the second-order scheme is now evaluated in the usual way, but with the left and right cell states ${\mathbf{U}_{i}^{n}}$ and ${\mathbf{U}_{i + 1}^{n}}$ replaced by the evolved boundary-extrapolated states ${\overline{\mathbf{U}_{i + \frac{1}{2}}^{L}}}$ and ${\overline{\mathbf{U}_{i + \frac{1}{2}}^{R}}}$, respectively, i.e. for the case of Lax-Friedrichs fluxes, one has:

\begin{equation}
\mathbf{F}_{i + \frac{1}{2}} = \frac{1}{2} \left[ \mathbf{F} \left( \overline{\mathbf{U}_{i + \frac{1}{2}}^{L}} \right) + \mathbf{F} \left( \overline{\mathbf{U}_{i + \frac{1}{2}}^{R}} \right) \right] - \frac{\Delta x}{2 \Delta t} \left[ \overline{\mathbf{U}_{i + \frac{1}{2}}^{R}} - \overline{\mathbf{U}_{i + \frac{1}{2}}^{L}} \right],
\end{equation}
and for the case of Roe fluxes, one has:

\begin{equation}
\mathbf{F}_{i + \frac{1}{2}} = \frac{1}{2} \left[ \mathbf{F} \left( \overline{\mathbf{U}_{i + \frac{1}{2}}^{L}} \right) + \mathbf{F} \left( \overline{\mathbf{U}_{i + \frac{1}{2}}^{R}} \right) \right] - \frac{1}{2} \sum_p \left\lvert \lambda_p \right\rvert \alpha_p \mathbf{r}_p,
\end{equation}
for Roe matrix ${\mathbf{A} \left( \overline{\mathbf{U}_{i + \frac{1}{2}}^{L}}, \overline{\mathbf{U}_{i + \frac{1}{2}}^{R}} \right)}$.

Flux limiters, and the resulting second-order extrapolations that they enable, satisfy the following properties:

\begin{theorem}
\label{th:flux1}
A flux limiter ${\phi \left( r \right)}$ will act symmetrically (i.e. will act equivalently on forward and backward gradients) if it satisfies the symmetry condition:

\begin{equation}
\frac{\phi \left( r \right)}{r} = \phi \left( \frac{1}{r} \right).
\end{equation}
\end{theorem}

\begin{theorem}
\label{th:flux2}
A second-order scheme extrapolated from an underlying first-order Lax-Friedrichs or Roe solver will be second-order TVD (total variation diminishing), i.e. one will have:

\begin{equation}
TV \left( \mathbf{U}^{n + 1} \right) \leq TV \left( \mathbf{U}^n \right),
\end{equation}
where ${TV \left( \mathbf{U}^n \right)}$ denotes the total variation of the solution at time ${t^n}$:

\begin{equation}
TV \left( \mathbf{U}^n \right) = \sum_i \left\lVert \mathbf{U}_{i + 1}^{n} - \mathbf{U}_{i}^{n} \right\rVert,
\end{equation}
if the flux limiter ${\phi \left( r \right)}$ satisfies the Sweby criteria\cite{sweby_high_1984}:

\begin{align}
\forall r < 0, \qquad & \phi \left( r \right) = 0\\
\forall 0 \leq r \leq \frac{1}{2}, \qquad & r \leq \phi \left( r \right) \leq 2r,\\
\forall \frac{1}{2} \leq r \leq 1, \qquad & r \leq \phi \left( r \right) \leq 1,\\
\forall 1 \leq r \leq 2, \qquad & 1 \leq \phi \left( r \right) \leq r,\\
\forall r > 2, \qquad & 1 \leq \phi \left( r \right) \leq 2,
\end{align}
with ${\phi \left( 1 \right) = 1}$.
\end{theorem}

In what follows, we shall focus upon four standard flux limiters in particular, all of which had previously been implemented as part of the \textsc{Gkeyll} code, namely the \textit{minmod} limiter\cite{roe_characteristic-based_1986}:

\begin{equation}
\phi_{mm} \left( r \right) = \max \left( 0, \min \left( 1, r \right) \right),
\end{equation}
the \textit{monotonized-centered} limiter\cite{van_leer_towards_1977}:

\begin{equation}
\phi_{mc} \left( r \right) = \max \left( 0, \min \left( 2r, \frac{1}{2} \left( 1 + r \right), 2 \right) \right),
\end{equation}
the \textit{superbee} limiter\cite{roe_characteristic-based_1986}:

\begin{equation}
\phi_{sb} \left( r \right) = \max \left( 0, \min \left( 2r, 1 \right), \min \left( r, 2 \right) \right),
\end{equation}
and the \textit{van Leer} limiter\cite{van_leer_towards_1974}:

\begin{equation}
\phi_{vl} \left( r \right) = \frac{r + \left\lvert r \right\rvert}{1 + \left\lvert r \right\rvert}.
\end{equation}

\section{Methodology and Results}
\label{sec:sec6}

\subsection{Automatic Code Generation}
\label{sec:sec7}

Before we proceed with formally verifying various properties of finite volume schemes in Racket, it is first necessary for us to be able to generate reliable C implementations which certifiably match the symbolic Racket expressions being reasoned about. To this end, we introduce a general data structure for representing hyperbolic PDE systems in Racket, consisting of four lists of symbolic Racket expressions: \lstinline{cons-exprs} representing the conserved variable vector, \lstinline{flux-exprs} representing the flux vector, \lstinline{max-speed-exprs} representing the maximum wave-speed estimates (used for enforcing the CFL stability condition), and \lstinline{parameters} representing any additional simulation parameters (such as equation of state variables). For example, the equations representing the density ${\rho}$ and $x$-momentum ${\rho u}$ components of the isothermal Euler equation system are represented in our Racket implementation as follows, assuming a thermal velocity ${v_{th} = 1.0}$:

\begin{lstlisting}
(define pde-system-isothermal-euler
  (hash
    'name "isothermal-euler"
    'cons-exprs (list `rho, `mom_x)
    'flux-exprs (list
      `mom_x
      `(+ (/ (* mom_x mom_x) rho)
        (* rho vt vt)))
    `max-speed-exprs (list
      `(abs (- (/ mom_x rho) vt))
      `(abs (+ (/ mom_x rho) vt)))
    `parameters (list `(define vt 1.0))))
\end{lstlisting}
Each of the Racket expressions in these lists may then be converted recursively into a string representing a functionally equivalent C expression, using the function \lstinline{convert-expr}:

\begin{lstlisting}
(define (convert-expr expr)
  (match expr
    [(? symbol? symb) (symbol->string symb)]
    [(? number? num) (number->string num)]
    ...
    [`(abs ,arg)
      (format "fabs(~a)" (convert-expr arg))]
    ...
    [`(+ . ,terms)
      (let ([c-terms
          (map convert-expr terms)])
        (string-append "("
          (string-join c-terms " + ") ")"))]
    ...))
\end{lstlisting}
The base case of \lstinline{convert-expr} converts any symbol or number directly into its corresponding string. Any single- or multi-argument function, e.g. \lstinline{(abs arg)} or \lstinline{(max arg1 arg2)}, is converted into its C equivalent, e.g. \lstinline[language=C]{fabs(arg)} or \lstinline[language=C]{fmax(arg1, arg2)}, with \lstinline{convert-expr} being called on all interior expressions. Finally, any elementary arithmetic operation, e.g. \lstinline{(+ arg1 arg2 ...)}, has \lstinline{convert-expr} mapped over each argument, and the results are interspersed with the corresponding arithmetic symbol, e.g. \lstinline[language=C]{+}, in C. We have generally erred on the side of over-generating parentheses in the resulting C code, in order to guarantee that the order of operations remains identical between the C and Racket versions of mathematical expressions. Conditional expressions in Racket are either converted into the corresponding ternary operators in C:

\begin{lstlisting}
(match expr
  [`(cond
      [,cond1 ,expr1]
      [else ,expr2])
    (format "(~a) ? ~a : ~a"
      (convert-expr cond1)
      (convert-expr expr1)
      (convert-expr expr2))])
\end{lstlisting}
or are converted directly into \lstinline[language=C]{if} statements. Finally, the strings generated by \lstinline{convert-expr} are spliced into a template for a generic finite volume solver using \lstinline{format}. Appropriate templates exist for both entirely standalone solvers, and for bespoke solver modules that can be integrated into the larger \textsc{Gkeyll} code.

\subsection{Symbolic Theorem-Proving}
\label{sec:sec8}

At its core, our automated theorem-proving framework is based upon a \textit{symbolic simplification algorithm}, which aims to reduce every symbolic Racket expression to some canonical algebraic form. Although such simplification algorithms are standard in computer algebra, our particular application makes this task non-trivial in two respects. First, we wish for our simplification algorithm to be based on a \textit{globally confluent}\cite{robinson_handbook_2001} and \textit{strongly normalizing}\cite{baader_term_1999} underlying rewriting system, since such rewriting systems exhibit highly desirable correctness and termination properties for the type of algebraic/equational theorem-proving with which we are concerned. This places certain restrictions on the kinds of rewriting rules that we are able to include, since many algebraically correct transformations, such as those corresponding to commutativity of addition or multiplication:

\begin{lstlisting}
(match expr
  [`(+ ,x ,y) (+ ,y ,x)]
  [`(* ,x ,y) (* ,y ,x)])
\end{lstlisting}
cannot be safely included without risking breaking the strong normalization property of the rewriting system (and hence termination of the theorem-prover). Second, since the variables over which we are reasoning typically represent floating-point numbers, and specifically \lstinline[language=C]{double}s in C, many standard algebraic rules, such as associativity of addition or multiplication:

\begin{lstlisting}
(match expr
  [`(+ ,x (+ ,y, z)) `(+ (+ ,x ,y) ,z)]
  [`(* ,x (* ,y ,z)) `(* (* ,x ,y) ,z)])
\end{lstlisting}
cannot safely be assumed to hold (except in certain restricted cases) due to the accumulation of truncation errors. For instance, in floating-point arithmetic\cite{goldberg_what_1991}:

\begin{equation}
\left( 10^{30} + -10^{30} \right) + 1 = 1, \qquad \text{ yet } \qquad 10^{30} + \left( -10^{30} + 1 \right) = 0.
\end{equation}
To this end, we have tried wherever possible to admit only to those algebraic transformations that are permitted under the IEEE 754 standard for floating-point arithmetic\cite{ieee_computer_society_ieee_2008}.

The particular collection of algebraic rewriting rules used within our theorem-prover is somewhat complex and ad hoc, so we shall only summarize the salient elements here. The basic structure consists of a recursively-defined \lstinline{symbolic-simp-rule} function of the form:

\begin{lstlisting}
(define (symbolic-simp-rule expr)
  (match expr
    [`(+ 0 ,x) `,x]
    [`(* 1 ,x) `,x]
    [`(* 0 ,x) 0]
    ...
    [`(+ . ,terms)
      `(+ ,@(map (lambda (term)
        (symbolic-simp-rule term)) terms))]
    ...
    [else expr]))
\end{lstlisting}
The first few rules shown above are examples of elementary algebraic properties\cite{blasius_deduktionssysteme:_1992}, such as the existence of 0 as a (left) additive identity, the existence of 1 as a (left) multiplicative identity, the existence of 0 as a (left) annihilator for multiplication, etc. The next rule is an example of how \lstinline{symbolic-simp-rule} is mapped over each term within an elementary arithmetic operation such as \lstinline{(+ arg1 arg2 ...)} (similar mappings are performed for operations such as \lstinline{(abs arg)} or \lstinline{(max arg1 arg2)}, with each subexpression being recursively simplified). Finally, if none of the rewriting rules match the expression, then the expression has reached a normal form and is returned verbatim. The symbolic simplifier itself then consists of a single function \lstinline{symbolic-simp} which recursively calls \lstinline{symbolic-simp-rule} until a fixed point is achieved (i.e. until the expression stops changing):

\begin{lstlisting}
(define (symbolic-simp expr)
  (define simp-expr (symbolic-simp-rule expr))
  (cond
    [(equal? simp-expr expr) expr]
    [else (symbolic-simp simp-expr)]))
\end{lstlisting}

In order to facilitate the reduction of all symbolic expressions to a canonical form, rewriting rules exist to move all numerical constants or coefficients to the left of non-numerical expressions within sums or products:

\begin{lstlisting}
(match expr
  [`(+ ,(and x (not (? number?)))
      ,(and y (? number?)))
    `(+ ,y ,x)]
  [`(* ,(and x (not (? number?)))
      ,(and y (? number?)))
    `(* ,y ,x)])
\end{lstlisting}
to collect ``like'' terms together within sums and differences (via the distributive property):

\begin{lstlisting}
(match expr
  [`(+ (* ,a ,x) (* ,b ,x)) `(* (+ ,a ,b) ,x)]
  [`(- (* ,a ,x) (* ,b ,x)) `(* (- ,a ,b) ,x)])
\end{lstlisting}
and to evaluate any arithmetic expressions or operations involving purely numerical values directly:

\begin{lstlisting}
(match expr
  [`(+ ,(and x (? number?))
    ,(and y (? number?))) (+ x y)]
  ...
  [`(sqrt ,(and x (? number?))) (sqrt x)]
  ...)
\end{lstlisting}
etc. A more complicated set of rules and heuristics exist regarding whether and when to expand brackets, factorize subexpressions, and so on. Algebraic rules are also defined for standard mathematical functions such as \lstinline{(sqrt ...)} or \lstinline{(abs ...)}, stating for instance that the square root of the square of a quantity, or the square of the square root of a quantity, is equal to the quantity itself:

\begin{lstlisting}
(match expr
  [`(sqrt (* ,x ,x)) `,x]
  [`(* (sqrt ,x) (sqrt ,x)) `,x])
\end{lstlisting}
or that the absolute value of the negation of a quantity is equal to the quantity itself:

\begin{lstlisting}
(match expr
  [`(abs (* -1 ,x)) `(abs ,x)])
\end{lstlisting}
or that the \lstinline{(max ...)} and \lstinline{(min ...)} functions satisfy (at least in the binary case):

\begin{equation}
\max \left( x, y \right) = \frac{1}{2} \left( x + y \right) + \frac{1}{2} \left\lvert x - y \right\rvert,
\end{equation}
and:

\begin{equation}
\min \left( x, y \right) = \frac{1}{2} \left( x + y \right) - \frac{1}{2} \left\lvert x - y \right\rvert,
\end{equation}
respectively, i.e:

\begin{lstlisting}
(match expr
  [`(max ,x ,y)
    `(+ (* 0.5 (+ ,x ,y))
      (* 0.5 (abs (- ,x ,y))))]
  [`(min ,x ,y)
    `(- (* 0.5 (+ ,x ,y))
      (* 0.5 (abs (- ,x ,y))))])
\end{lstlisting}
etc.

\subsection{Automatic Differentiation}
\label{sec:sec9}

In order to compute symbolic derivatives of arbitrary Racket expressions, we implement a minimalistic \textit{automatic differentiation} algorithm, again restricting ourselves to assume only those algebraic properties which hold for arbitrary floating-point numbers. Any Racket expression \lstinline{expr} may then be differentiated symbolically with respect to the variable \lstinline{var}, by means of the function \lstinline{symbolic-diff}:

\begin{lstlisting}
(define (symbolic-diff expr var)
  (match expr
    [(? symbol? symb) (cond
      [(eq? symb var) 1.0]
      [else 0.0])]
    [(? number?) 0.0]
    ...
    [`(+ . ,terms)
      (+ ,@(map (lambda (term)
        (symbolic-diff term var)) terms))]
    ...))
\end{lstlisting}
The base case of \lstinline{symbolic-diff} evaluates the derivative of any symbol or numerical constant to 0, unless that symbol matches the variable with respect to which one is differentiating, in which case it evaluates to 1. For any sum of the form \lstinline{(+ arg1 arg2 ...)}, \lstinline{symbolic-diff} is mapped over each term, reflecting the linearity of differentiation (and likewise for differences). For any product of the form \lstinline{(* arg1 arg2 ...)}, \lstinline{symbolic-diff} is applied to each term in the product separately (with all other terms being kept fixed), with the results then being summed together:

\begin{lstlisting}
(match expr
  [`(* . ,terms)
    ((lambda (sums) (cond
        [(null? (cdr sums)) (car sums)]
        [else (cons '+ sums)]))
      (let loop ([i ])
        (cond
          [(= i (length terms)) `()]
          [else (let ([di (symbolic-diff
            (list-ref terms i) var)])
              (cons (cons '* (for/list
                [(j (in-range (length terms))])
                  (cond
                    [(= j i) di]
                    [else (list-ref terms j)])))
              (loop (add1 i))))])))])
\end{lstlisting}
reflecting the product rule of differentiation, etc. Certain standard mathematical functions, such as \lstinline{(sqrt ...)}, also have their derivatives specifically encoded wherever they are well-defined:

\begin{lstlisting}
(match expr
  [`(sqrt ,x) `(* 0.5 (/ 1.0 (sqrt ,x)))])
\end{lstlisting}

With the ability to differentiate arbitrary scalar functions thus in place, the symbolic Jacobian of a list of symbolic Racket expressions \lstinline{exprs}, evaluated with respect to a list of symbolic Racket variables \lstinline{vars}, may now be computed via a straightforward \lstinline{symbolic-jacobian} function:

\begin{lstlisting}
(define (symbolic-jacobian exprs vars)
  (map (lambda (expr)
    (map (lambda (var)
      (symbolic-simp (symbolic-diff expr var)))
    vars
  exprs))
\end{lstlisting}
Likewise for the symbolic gradient of a single symbolic Racket expression \lstinline{expr}, with respect to a list of symbolic Racket variables \lstinline{vars}, via the \lstinline{symbolic-gradient} function:

\begin{lstlisting}
(define (symbolic-gradient expr vars)
  (map (lambda (var)
    (symbolic-simp (symbolic-diff expr var)))
  vars))
\end{lstlisting}
The symbolic Hessian (\lstinline{symbolic-hessian}) of a symbolic Racket expression may therefore be computed as a composition of the two:

\begin{lstlisting}
(define (symbolic-hessian expr vars)
  (symbolic-jacobian (symbolic-gradient expr
    vars) vars))
\end{lstlisting}

\subsection{Stability, Hyperbolicity and Convexity}
\label{sec:sec10}

In order to prove that a given Lax-Friedrichs solver meets the CFL stability criterion listed in Theorem \ref{th:lax2}, it is sufficient to compute the flux Jacobian ${\mathbf{J}_{\mathbf{F}}}$ by evaluating:

\begin{lstlisting}
(symbolic-jacobian flux-exprs cons-vars)
\end{lstlisting}
and then to compute its symbolic eigenvalues, which for instance can be done in the case of a 2x2 matrix using:

\begin{lstlisting}
(define (symbolic-eigvals2 matrix)
  (let ([a (list-ref (list-ref matrix 0) 0)]
    [b (list-ref (list-ref matrix 0) 1)]
    [c (list-ref (list-ref matrix 1) 0)]
    [d (list-ref (list-ref matrix 1) 1)])
    (list
      `(* 0.5 (+ (- ,a (sqrt (+ (* 4.0 ,b ,c)
        (* (- ,a ,d) (- ,a ,d))))) ,d))
      `(* 0.5 (+ (+ ,a (sqrt (+ (* 4.0 ,b ,c)
        (* (- ,a ,d) (- ,a ,d))))) ,d)))])))
\end{lstlisting}
where we have simply encoded the explicit solutions ${\lambda_{\pm}}$ of the characteristic polynomial for the 2x2 matrix:

\begin{equation}
M = \begin{bmatrix}
a & b\\
c & d
\end{bmatrix},
\end{equation}
i.e:

\begin{equation}
\lambda_{\pm} = \frac{1}{2} \left( a + d \pm \sqrt{a^2 + 4 b c - 2 a d + d^2} \right).
\end{equation}
We can now proceed to apply \lstinline{(abs ...)} to each symbolic eigenvalue, and then compare the absolute eigenvalues against the expressions in \lstinline{max-speed-exprs} (after mapping \lstinline{symbolic-simp} over the two lists of expressions, reducing them to their respective normal forms) in order to confirm that they are indeed identical. Since the time-steps ${\Delta t}$ within the generated C code are computed directly from the elements of \lstinline{max-speed-exprs} as:

\begin{equation}
\Delta t = \frac{C_{CFL} \Delta x}{\left\lvert a \right\rvert},
\end{equation}
where ${\left\lvert a \right\rvert}$ is the largest value in \lstinline{max-speed-exprs}, this condition is sufficient to guarantee CFL stability under Theorem \ref{th:lax2} provided that ${0 < C_{CFL} \leq 1}$, which is also verified by the theorem-prover during the initialization step:

\begin{lstlisting}
(cond
  [(or (<= cfl 0) (> cfl 1)) #f]
  ...
  [else #t])
\end{lstlisting}

In order to prove that a given Lax-Friedrichs solver meets the hyperbolicity-preservation criterion listed in Theorem 
\ref{th:lax1}, one must, furthermore, determine whether each of the symbolic eigenvalues of the flux Jacobian ${\mathbf{J}_{\mathbf{F}}}$ corresponds to a real number. For this purpose, we introduce an \lstinline{is-real} function:

\begin{lstlisting}
(define (is-real expr cons-var parameters)
  (match expr
    [(? real?) #t]
    ...
    [(? (lambda (arg)
      (not (equal? (member arg cons-vars)
        #f)))) #t]
    ...
    [(? (lambda (arg)
      (and (not (empty? parameters)) (ormap
        (lambda (parameter) (equal? arg
          (list-ref parameter 1)))
        parameters)))) #t]
    ...
    [(else #f)]))
\end{lstlisting}
The base case of \lstinline{is-real} treats any real numerical constant as a real number, and any simulation parameter or conserved variable is also assumed to be real by default. This latter condition is then enforced during the simulation initialization step itself:

\begin{lstlisting}
(cond
  [(not (or (empty? parameters) (andmap
    (lambda (parameter) (is-real
      (list-ref parameter 2) (list cons-expr)
      parameters)) parameters))) #f]
  [(not (is-real init-func (list cons-expr)
    parameters)) #f]
  ...
  [else #t])
\end{lstlisting}
We enforce that the set of real numbers is closed under operations like addition, subtraction, and multiplication, e.g:

\begin{lstlisting}
(match expr
  [`(+ . ,terms)
    (andmap (lambda (term)
      (is-real term cons-vars parameters))
    terms)]
  ...)
\end{lstlisting}
etc., and also that the set of reals remains closed under division so long as the denominator is non-zero, and under square roots so long as the argument is non-negative; these latter two conditions are enforced via the additional functions \lstinline{is-non-zero} and \lstinline{is-non-negative}, which will be described momentarily.

Moreover, \textit{strict} hyperbolicity-preservation of a given Lax-Friedrichs solver can be proven (in the 2x2 case) by proving that the symbolic eigenvalues of the flux Jacobian ${\mathbf{J}_{\mathbf{F}}}$ are not only real but also distinct, using the \lstinline{are-distinct} function:

\begin{lstlisting}
(define (are-distinct expr parameters)
  (match expr
    [(? (lambda (arg) (and (number?
      (list-ref arg 0)) (number?
        (list-ref arg 1))
      (not (equal? (list-ref arg 0)
        (list-ref arg 1)))))) #t]
  ...
  [else #f]))
\end{lstlisting}
The base case of \lstinline{are-distinct} treats any pair of non-equal numerical constants as distinct. Pairs of expressions of the form ${\left\lbrace x, -x \right\rbrace}$ or ${\left\lbrace x + y, x - y \right\rbrace}$ are then also treated as distinct, provided that the expressions $x$ or $y$ are themselves non-zero, respectively:

\begin{lstlisting}
(match expr
  [`(,x (* -1 ,x)) (is-non-zero x parameters)]
  [`((* -1 ,x) ,x) (is-non-zero x parameters)]  
  [`((+ ,x ,y) (- ,x ,y))
    (is-non-zero y parameters)]
  [`((- ,x ,y) (+ ,x ,y))
    (is-non-zero y parameters)]
  ...)
\end{lstlisting}
where the function \lstinline{is-non-zero} treats all non-zero numerical constants as non-zero, as its base case:

\begin{lstlisting}
(define (is-non-zero expr parameters)
  (match expr
    [(? lambda (arg)
      (and (number? arg) (not (equal? arg 0)))))
        #t]
    ...
    [else #f]))
\end{lstlisting}
It also treats all simulation parameters as non-zero, and enforces this condition, using similar logic to what was previously described for \lstinline{is-real}. Finally, it treats the product of any two non-zero expressions as non-zero:

\begin{lstlisting}
(match expr
  [`(* ,x ,y) (and (is-non-zero x parameters)
    (is-non-zero y parameters))])
\end{lstlisting}

Finally, in order to prove that a given Lax-Friedrichs solver meets the local Lipschitz continuity criterion listed in Theorems \ref{th:lax3} and \ref{th:lax4}, it is sufficient to compute the symbolic Hessian ${\mathbf{H}_f}$ by evaluating:

\begin{lstlisting}
(symbolic-hessian flux-expr cons-exprs)
\end{lstlisting}
for each \lstinline{flux-expr} $f$ in the list \lstinline{flux-exprs}, and then to compute its symbolic eigenvalues using \lstinline{symbolic-eigvals2} (at least in the 2x2 case), in order to confirm that each symbolic eigenvalue is non-negative. Non-negativity can be checked using the \lstinline{is-non-negative} function:

\begin{lstlisting}
(define (is-non-negative expr parameters)
  (match expr
    [(? lambda (arg)
      (and (number? arg) (>= arg 0)))) #t]
    ...
    [else #f]))
\end{lstlisting}
whose base case treats all non-negative numerical constants as non-negative. Moreover, we enforce that the sum, product or quotient of two non-negative numbers is always non-negative, e.g:

\begin{lstlisting}
(match expr
  [`(* ,x ,y) (and
    (is-non-negative x parameters)
    (is-non-negative y parameters))]
  ...)
\end{lstlisting}
etc. Although we have described these techniques in the context of proving the hyperbolicity-preserving and strict hyperbolicity-preserving properties of Lax-Friedrichs solvers via Theorem \ref{th:lax1}, we note that the same techniques can be used to prove the hyperbolicity-preserving and strict hyperbolicity-preserving properties of Roe solvers also, via Theorem \ref{th:roe1}, by exploiting the fact that a valid symbolic Roe matrix ${\mathbf{A} \left( \mathbf{U}_{i}^{n}, \mathbf{U}_{i + 1}^{n} \right)}$ can be calculated as an average of the two symbolic flux Jacobians:

\begin{equation}
\mathbf{A} \left( \mathbf{U}_{i}^{n}, \mathbf{U}_{i + 1}^{n} \right) = \frac{1}{2} \left[ \nabla_{\mathbf{U}} \mathbf{F} \left( \mathbf{U}_{i}^{n} \right) + \nabla_{\mathbf{U}} \mathbf{F} \left( \mathbf{U}_{i + 1}^{n} \right) \right],
\end{equation}
and the reality and distinctness of its eigenvalues can therefore be verified in exactly the same way as for ${\mathbf{J}_{\mathbf{F}}}$. The flux conservation/jump continuity condition can be verified purely algebraically, by simply confirming that the two sides of the equation in Theorem \ref{th:roe2} both reduce to the same canonical form using \lstinline{symbolic-simp}.

\subsection{Symbolic Limits and Symmetry}
\label{sec:sec11}

In order to determine whether a given flux limiter ${\phi \left( r \right)}$ satisfies the symmetry condition presented in Theorem \ref{th:flux1}, it is necessary to perform a ${r \to \frac{1}{r}}$ variable transformation throughout the expression for ${\phi \left( r \right)}$, in order to determine whether this transformed expression is algebraically equivalent to ${\frac{\phi \left( r \right)}{r}}$. For this purpose, we use the recursively-defined \lstinline{variable-transform} function:

\begin{lstlisting}
(define (variable-transform expr var new-var)
  (cond
    [(symbol? expr) (cond
      [(equal? expr var) new-var]
      [else expr])]
    [(pair? expr] (map (lambda (subexpr)
      (variable-transform subexpr var new-var))
      expr)]
    [else expr]))
\end{lstlisting}
which applies itself recursively to all subexpressions, and replaces any subexpression it finds which matches \lstinline{var} with a new subexpression matching \lstinline{new-var}. For the second-order total variation diminishing (TVD) condition presented in Theorem \ref{th:flux2}, we use a strictly stronger set of conditions which imply (but are not equivalent to) the Sweby criteria, namely the limiting conditions that:

\begin{equation}
0 \leq \lim_{r \to 0} \left[ \phi \left( r \right) \right] \leq 1, \qquad 1 \leq \lim_{r \to 2} \left[ \phi \left( r \right) \right] \leq 2,
\end{equation}
\begin{equation}
\lim_{r \to 1} \left[ \phi \left( r \right) \right] = 1, \qquad \lim_{r \to \infty} \left[ \phi \left( r \right) \right] \leq 2,
\end{equation}
assuming that ${\phi \left( r \right) = 0}$ for all ${r < 0}$, combined with the condition that ${\phi \left( r \right)}$ be a (non-strictly) concave function:

\begin{equation}
\frac{d^2 \phi \left( r \right)}{d r^2} \leq 0.
\end{equation}
These symbolic limits of the form ${\lim\limits_{x \to x_0} \left[ f \left( x \right) \right]}$ are evaluated via the \lstinline{evaluate-limit} function, which first performs a variable transformation ${x \to x_0}$ (using the extended reals ${x \in \mathbb{R} \cup \left\lbrace -\infty, +\infty \right\rbrace}$), and then recursively simplifies until the limiting expression achieves a fixed point:

\begin{lstlisting}
(define (evaluate-limit expr var limit)
  (define limit-val
    (variable-transform expr var limit))
  (define limit-expr
    (evaluate-limit-rule limit-val var limit))
  (cond
    [(equal? limit-expr expr) expr]
    [else (evaluate-limit limit-expr var limit)]))
\end{lstlisting}
where \lstinline{evaluate-limit-rule} encodes valid algebraic simplification rules over the extended reals (a specialized subset of the algebraic rules encoded by \lstinline{symbolic-simp-rule}).

\subsection{Results}
\label{sec:sec12}

For the two scalar PDEs (i.e. the linear advection and inviscid Burgers' equations), the hyperbolicity-preservation, CFL stability and local Lipschitz continuity properties of the Lax-Friedrichs solver, and the hyperbolicity-preservation and flux conservation (jump continuity) properties of the Roe solver, can all be proved directly and without further complication, as outlined in Tables \ref{tab:scalar1} and \ref{tab:scalar2}. For the two vector PDE systems (i.e. the perfectly hyperbolic Maxwell's equations and isothermal Euler equations), slightly more work is required. In order to ensure that the theorem-prover never needs to manipulate any higher-dimensional (i.e. beyond 2x2) matrices, we first ``factorize'' these equation systems into coupled pairs of PDEs, with each pair being dealt with independently. Maxwell's equations effectively ``factorize'' into four coupled pairs of linear advection equations, for the ${E^y}$ and ${B^z}$ components, the ${E^z}$ and ${B^y}$ components, the ${E^x}$ and ${\phi}$ components, and the ${B^x}$ and ${\psi}$ components, respectively. The isothermal Euler equations effectively ``factorize'' into a non-linear pair of equations for the ${\rho}$ and ${\rho u}$ components, and a linear pair of equations for the ${\rho v}$ and ${\rho w}$ components (since the ${\rho v}$ and ${\rho w}$ momentum components can each be regarded as being advected linearly with a speed of $u$ at each time-step). As shown in Tables \ref{tab:vector1} and \ref{tab:vector2}, for all four coupled systems constituting the perfectly hyperbolic Maxwell's equations, the hyperbolicity-preservation, strict hyperbolicity-preservation, CFL stability and local Lipschitz continuity properties of the Lax-Friedrichs solver, and the hyperbolicity-preservation, strict hyperbolicity-preservation and flux conservation (jump continuity) properties of the Roe solver, can all be proved unproblematically. However, for the two coupled systems constituting the isothermal Euler equations, several properties cannot immediately be proven: for the Lax-Friedrichs solver, proofs cannot be found for the local Lipschitz continuity property for the ${\rho}$ and ${\rho u}$ components, or for the strict hyperbolicity-preservation property for the ${\rho v}$ and ${\rho w}$ components, while for the Roe solver, proofs cannot be found for \textit{any} of the properties for the ${\rho}$ and ${\rho u}$ components, or for the strict hyperbolicity-preservation property for the ${\rho v}$ and ${\rho w}$ components. It is worth expanding upon each of these ``failure'' cases in greater detail.

\begin{table*}[ht]
\centering
\begin{tabular}{|c|c|c|c|c|c|}
\hline
\textbf{Equation} & \textbf{Hyperbolicity (Lax)} & \textbf{Stability (Lax)} & \textbf{Local Lipschitz (Lax)}\\
\hline
Linear Advection & 33 & 45 & 39\\
\hline
Inviscid Burgers' & 90 & 116 & 96\\
\hline
\end{tabular}
\caption{Numbers of proof steps required to prove hyperbolicity-preservation, CFL stability and local Lipschitz continuity for the Lax-Friedrichs solver, across both the linear advection and inviscid Burgers' scalar conservation equations.}
\label{tab:scalar1}
\end{table*}

\begin{table*}[ht]
\centering
\begin{tabular}{|c|c|c|}
\hline
\textbf{Equation} & \textbf{Hyperbolicity (Roe)} & \textbf{Conservation (Roe)}\\
\hline
Linear Advection & 55 & 92\\
\hline
Inviscid Burgers' & 120 & 188\\
\hline
\end{tabular}
\caption{Numbers of proof steps required to prove hyperbolicity-preservation and flux conservation (jump continuity) for the Roe solver, across both the linear advection and inviscid Burgers' scalar conservation equations.}
\label{tab:scalar2}
\end{table*}

\begin{table*}[ht]
\centering
\begin{tabular}{|c|c|c|c|c|}
\hline
\textbf{Equations} & \textbf{Hyperbolicity (Lax)} & \textbf{Strict Hyperbolicity (Lax)} & \textbf{Stability (Lax)} & \textbf{Local Lipschitz (Lax)}\\
\hline
Maxwell's (${E^y}$ and ${B^z}$) & 499 & 502 & 575 & 272\\
\hline
Maxwell's (${E^z}$ and ${B^y}$) & 627 & 630 & 711 & 452\\
\hline
Maxwell's (${E^x}$ and ${\phi}$) & 729 & 735 & 803 & 450\\
\hline
Maxwell's (${B^x}$ and ${\psi}$) & 731 & 737 & 805 & 450\\
\hline
Isothermal Euler (${\rho}$ and ${\rho u}$) & 1462 & 1465 & 1558 & -\\
\hline
Isothermal Euler (${\rho v}$ and ${\rho w}$) & 1189 & - & 1329 & 251\\
\hline
\end{tabular}
\caption{Numbers of proof steps (where applicable) required to prove hyperbolicity-preservation, strict hyperbolicity-preservation, CFL stability and local Lipschitz continuity for the Lax-Friedrichs solver, across all coupled pairs of components for the perfectly hyperbolic Maxwell's and isothermal Euler equation systems.}
\label{tab:vector1}
\end{table*}

\begin{table*}
\centering
\begin{tabular}{|c|c|c|c|}
\hline
\textbf{Equations} & \textbf{Hyperbolicity (Roe)} & \textbf{Strict Hyperbolicity (Roe)} & \textbf{Conservation (Roe)}\\
\hline
Maxwell's (${E^y}$ and ${B^z}$) & 616 & 619 & 412\\
\hline
Maxwell's (${E^z}$ and ${B^y}$) & 826 & 829 & 643\\
\hline
Maxwell's (${E^x}$ and ${\phi}$) & 871 & 877 & 642\\
\hline
Maxwell's (${B^x}$ and ${\psi}$) & 873 & 879 & 642\\
\hline
Isothermal Euler (${\rho}$ and ${\rho u}$) & - & - & -\\
\hline
Isothermal Euler (${\rho v}$ and ${\rho w}$) & 1315 & - & 498\\
\hline
\end{tabular}
\caption{Numbers of proof steps (where applicable) required to prove hyperbolicity-preservation, strict hyperbolicity-preservation and flux conservation (jump continuity) for the Roe solver, across all coupled pairs of components for the perfectly hyperbolic Maxwell's and isothermal Euler equation systems.}
\label{tab:vector2}
\end{table*}

\begin{table*}
\centering
\begin{tabular}{|c|c|c|}
\hline
\textbf{Flux Limiter} & \textbf{Symmetry} & \textbf{Second-Order TVD}\\
\hline
minmod & 863 & 513\\
\hline
monotonized-centered & 4171 & 2251\\
\hline
superbee & - & 2125\\
\hline
van Leer & 244 & -\\
\hline
\end{tabular}
\caption{Numbers of proof steps (where applicable) required to prove the symmetry and second-order total variation diminishing properties of the minmod, monotonized-centered, superbee and van Leer flux limiters.}
\label{tab:limiters}
\end{table*}

By interrogating the attempted proof of local Lipschitz continuity for the Lax-Friedrichs solver for the ${\rho}$ and ${\rho u}$ components, we see that the theorem-prover succeeds in reducing the problem of proving flux convexity to the problem of proving that the inequality:

\begin{equation}
\frac{2 \left( \left( \rho u \right)^2 + \rho^2 \right)}{\rho^3} \geq 0,
\end{equation}
always holds. However, in the absence of any guarantee that ${\rho > 0}$, it is unable to proceed further. Thus, we see that the theorem-prover has correctly concluded that it is unable to guarantee local Lipschitz continuity of the flux function in the absence of the additional constraint that the fluid density ${\rho}$ always be strictly positive (which the solver in isolation does not guarantee). The theorem-prover is also correct to conclude that strict hyperbolicity-preservation for the Lax-Friedrichs solver for the ${\rho v}$ and ${\rho w}$ components does not hold, due to the presence of a repeated ``$u$'' eigenvalue within the flux Jacobian ${\mathbf{J}_{\mathbf{F}}}$. Likewise, by interrogating the attempted proofs of hyperbolicity- and strict hyperbolicity-preservation for the Roe solver for the ${\rho}$ and ${\rho u}$ components, we see that the theorem-prover reduces these problems to the problem of proving that the quantities:

\begin{multline}
\lambda_{\pm} = \frac{1}{2 \rho_{L}^{2} \rho_{R}^{2}} \left[ \rho_L \rho_R \left( \left( \rho_R u_R \right) \rho_L + \left( \rho_L u_L \right) \rho_R \right) \right]\\
\pm \frac{1}{2 \rho_{L}^{2} \rho_{R}^{2}} \left[ \sqrt{- \rho_{L}^{2} \rho_{R}^{2} \left( \left( \rho_R u_R \right) \rho_L - \left( \rho_L u_L \right) \rho_R \right)^2 + \rho_{"L}^{4} \rho_{R}^{4} v_{th}^{2}} \right],
\end{multline}
are always real and distinct, which itself can only be true if the inequality:

\begin{equation}
4 \rho_{L}^{4} \rho_{R}^{4} v_{th}^{2} \geq \rho_{L}^{2} \rho_{R}^{2} \left( \left( \rho_R u_R \right) \rho_L - \left( \rho_L u_L \right) \rho_R \right)^2,
\end{equation}
is satisfied, where ${\rho_L}$, ${\rho_L u_L}$ and ${\rho_R}$, ${\rho_R u_R}$ denote the fluid density ${\rho}$ and fluid momentum ${\rho u}$ within the left and right cells ${x_i}$ and ${x_{i + 1}}$, respectively. Therefore, once again, we find that the theorem-prover correctly determines that these properties cannot be guaranteed in the absence of some form of positivity restriction on the fluid density ${\rho}$. Similarly, the presence of the repeated ``$u$'' eigenvalue within the Roe matrix ${\mathbf{A} \left( \mathbf{U}_{i}^{n}, \mathbf{U}_{i + 1}^{n} \right)}$ for the ${\rho v}$ and ${\rho w}$ components implies that strict hyperbolicity-preservation does not hold here either. Hence, we see that the only property for which the theorem-prover truly \textit{fails} (i.e. where it does not succeed in finding a proof for a statement which is unconditionally true) is the flux conservation/jump continuity condition for the Roe solver for the ${\rho}$ and ${\rho u}$ components.

For the minmod and monotonized-centered flux limiters, the symmetry and second-order TVD properties can both be proved directly and unproblematically, as shown in Table \ref{tab:limiters}. However, we see that the theorem-prover fails to find valid proofs for the symmetry property of the superbee limiter, and the second-order TVD property of the van Leer limiter. As in the case of the flux conservation condition for the isothermal Euler Roe solver, this is due entirely to present algebraic limitations of the simplification algorithm (since these properties certainly do hold, unconditionally, in the case of these limiters); these limitations can presumably be circumvented via the judicious introduction of stronger symbolic simplification rules into \lstinline{symbolic-simp-rule} and related functions.

\section{Conclusions and Future Work}
\label{sec:sec13}

In this paper, we have introduced a new formal verification pipeline in Racket for first-order hyperbolic PDE solvers, with a particular emphasis upon finite volume, high-resolution shock-capturing methods. Although the resulting automated theorem-proving framework still exhibits some notable limitations, we see that it is nevertheless able to produce full proofs of correctness for several linear and non-linear hyperbolic PDE solvers (with both first- and second-order spatial accuracy), and to produce conditional/partial proofs of correctness for others. These general correctness results encompass both mathematical (e.g. hyperbolicity-preservation) and physical (e.g. thermodynamic validity) notions of correctness. At present, this pipeline constitutes around 15,000 lines of Racket in total, although the vast majority of this is boilerplate for the purposes of synthesizing functionally complete C code (both in the standalone case, and for integration into the \textsc{Gkeyll} codebase, which includes automatic synthesis of header files, regression tests, etc.): the core theorem-proving, automatic differentiation, and symbolic limit evaluation routines fit within just a few hundred lines of Racket each, and it is likely that these can all be optimized further. We consider this work to be a successful proof-of-concept, representing a solid foundation upon which further such verification pipelines may be built, expanding into a wider variety of numerical solvers, equation systems, reconstruction algorithms, and discretization schemes.

In addition to reinforcing the weak points in the existing theorem-prover, such as adding new and more powerful rewriting rules (for instance, those needed to complete the proofs of correctness for the superbee and van Leer flux limiters) and introducing more systematic techniques for proving conditional results (such as those needed to complete the formalization of the conditional correctness proofs for the isothermal Euler solvers), we intend to expand this overall framework in several key directions. One such direction involves introducing formal verification tools for ordinary differential equation (ODE) integrators too, including the kinds of both explicit (e.g. strong stability-preserving Runge-Kutta\cite{gottlieb_strong_2001}) and implicit (e.g. time-centered Crank-Nicolson) ODE integrators used within the \textsc{Gkeyll} code, for instance when coupling hydrodynamic and electromagnetic PDE systems together\cite{wang_exact_2020} or handling geometric source terms in general relativity\cite{gorard_tetrad_2024}. Being able to automate the deduction of properties such as absolute stability regions for explicit Runge-Kutta integrators would, in itself, be an important and highly useful advance. Another major frontier of expansion would be the formalization of the kinds of modal discontinuous Galerkin (DG) algorithms used for solving kinetic equations within codes such as \textsc{Gkeyll}\cite{juno_discontinuous_2018}, for which properties like ${L^2}$ stability and entropy convexity are significantly more subtle and fragile than they are in the case of finite volume solvers, depending sensitively upon particular choices of basis functions\cite{juno_2020}, etc. Overall, it is hoped that the construction of \textit{formally verified simulations}, of the kind described here, will become an increasingly pervasive methodology throughout scientific computing in general, and within computational physics in particular.

\clearpage

\begin{acks}
J.G. was partially funded by the Princeton University Research Computing group. J.G. and A.H. were partially funded by the U.S. Department of Energy under Contract No. DE-AC02-09CH1146 via an LDRD grant. The development of \textsc{Gkeyll} was partially funded by the NSF-CSSI program, Award Number 2209471. J.G. thanks Nikola Bukowiecka for her proofreading and suggestions.
\end{acks}

\bibliographystyle{ACM-Reference-Format}
\bibliography{provablepaper-bib}

\end{document}

%% file: listings-scheme.tex
\definecolor{scheme-primitive}{rgb}{0, 0, 1}
\definecolor{scheme-derived}{rgb}{0, 0, 1}
\definecolor{scheme-comment}{rgb}{0, 0.5, 0}
\definecolor{scheme-bracket}{rgb}{0.6, 0.6, 0.6}

\lstdefinelanguage{Scheme}{
    basicstyle=\small\ttfamily,
    sensitive=true,
    morestring = [b]",
    stringstyle=\color{red},
    morecomment=[l]{;},
    commentstyle=\color{scheme-comment},
    literate=
            *{(}{{\textcolor{scheme-bracket}{(}}}{1}
            {)}{{\textcolor{scheme-bracket}{)}}}{1},
    classoffset=0,
    morekeywords={
        lambda,
        if,
        quote,
        set!},
    keywordstyle=\color{scheme-primitive},
    classoffset=1,
    morekeywords={
        cond,
        else,
        case,
        and,
        or,
        when,
        unless,
        cond-expand,
        let,
        let*,
        letrec,
        letrec*
        let-values,
        let*-values,
        begin,
        do,
        delay,
        delay-force,
        force,
        promise?,
        make-promise,
        make-parameter,
        parameterize,
        guard,
        quasiquote,
        case-lambda,
        let-syntax,
        letrec-syntax,
        syntax-rules,
        define,
        match,
        map},
    keywordstyle=\color{scheme-derived},
    classoffset=0
}

%% file: provablepaper.bbl

\begin{thebibliography}{37}

%%% the Web address.

\ifx \showCODEN    \undefined \def \showCODEN     #1{\unskip}     \fi
\ifx \showISBNx    \undefined \def \showISBNx     #1{\unskip}     \fi
\ifx \showISBNxiii \undefined \def \showISBNxiii  #1{\unskip}     \fi
\ifx \showISSN     \undefined \def \showISSN      #1{\unskip}     \fi
\ifx \showLCCN     \undefined \def \showLCCN      #1{\unskip}     \fi
\ifx \shownote     \undefined \def \shownote      #1{#1}          \fi
\ifx \showarticletitle \undefined \def \showarticletitle #1{#1}   \fi
\ifx \showURL      \undefined \def \showURL       {\relax}        \fi
% invisible to TeX
\providecommand\bibfield[2]{#2}
\providecommand\bibinfo[2]{#2}
\providecommand\natexlab[1]{#1}
\providecommand\showeprint[2][]{arXiv:#2}

\bibitem[Auger and Plagnol(2017)]%
        {auger_overview_2017}
\bibfield{author}{\bibinfo{person}{Gerard Auger} {and} \bibinfo{person}{Eric
  Plagnol}.} \bibinfo{year}{2017}\natexlab{}.
\newblock \bibinfo{booktitle}{\emph{An overview of gravitational waves: theory,
  sources and detection}}.
\newblock \bibinfo{publisher}{World scientific}, \bibinfo{address}{New Jersey}.
\newblock
\showISBNx{9789813141759}


\bibitem[Baader and Nipkow(1999)]%
        {baader_term_1999}
\bibfield{author}{\bibinfo{person}{Franz Baader} {and} \bibinfo{person}{Tobias
  Nipkow}.} \bibinfo{year}{1999}\natexlab{}.
\newblock \bibinfo{booktitle}{\emph{Term rewriting and all that}
  (\bibinfo{edition}{1st paperback edition} ed.)}.
\newblock \bibinfo{publisher}{Cambridge University Press},
  \bibinfo{address}{Cambridge New York Melbourne Madrid Cape Town}.
\newblock
\showISBNx{9780521779203 9780521455206}


\bibitem[Bläsius and Bürckert(1992)]%
        {blasius_deduktionssysteme:_1992}
\bibfield{editor}{\bibinfo{person}{Karl~Hans Bläsius} {and}
  \bibinfo{person}{Hans-Jürgen Bürckert}} (Eds.).
  \bibinfo{year}{1992}\natexlab{}.
\newblock \bibinfo{booktitle}{\emph{Deduktionssysteme: {Automatisierung} des
  logischen {Denkens}} (\bibinfo{edition}{2., völlig überarbeitete und
  erweiterte auflage} ed.)}.
\newblock \bibinfo{publisher}{Oldenbourg}, \bibinfo{address}{München Wien}.
\newblock
\showISBNx{9783486220339}


\bibitem[Breuß(2004)]%
        {breus_correct_2004}
\bibfield{author}{\bibinfo{person}{Michael Breuß}.}
  \bibinfo{year}{2004}\natexlab{}.
\newblock \showarticletitle{The correct use of the {Lax}–{Friedrichs}
  method}.
\newblock \bibinfo{journal}{\emph{ESAIM: Mathematical Modelling and Numerical
  Analysis}} \bibinfo{volume}{38}, \bibinfo{number}{3} (\bibinfo{date}{May}
  \bibinfo{year}{2004}), \bibinfo{pages}{519--540}.
\newblock
\showISSN{0764-583X, 1290-3841}
\href{https://doi.org/10.1051/m2an:2004027}{doi:\nolinkurl{10.1051/m2an:2004027}}


\bibitem[Courant et~al\mbox{.}(1967)]%
        {courant_partial_1967}
\bibfield{author}{\bibinfo{person}{R. Courant}, \bibinfo{person}{K.
  Friedrichs}, {and} \bibinfo{person}{H. Lewy}.}
  \bibinfo{year}{1967}\natexlab{}.
\newblock \showarticletitle{On the {Partial} {Difference} {Equations} of
  {Mathematical} {Physics}}.
\newblock \bibinfo{journal}{\emph{IBM Journal of Research and Development}}
  \bibinfo{volume}{11}, \bibinfo{number}{2} (\bibinfo{date}{March}
  \bibinfo{year}{1967}), \bibinfo{pages}{215--234}.
\newblock
\showISSN{0018-8646, 0018-8646}
\href{https://doi.org/10.1147/rd.112.0215}{doi:\nolinkurl{10.1147/rd.112.0215}}


\bibitem[Felleisen et~al\mbox{.}(2015)]%
        {felleisen_et_al}
\bibfield{author}{\bibinfo{person}{Matthias Felleisen},
  \bibinfo{person}{Robert~Bruce Findler}, \bibinfo{person}{Matthew Flatt},
  \bibinfo{person}{Shriram Krishnamurthi}, \bibinfo{person}{Eli Barzilay},
  \bibinfo{person}{Jay McCarthy}, {and} \bibinfo{person}{Sam Tobin-Hochstadt}.}
  \bibinfo{year}{2015}\natexlab{}.
\newblock \showarticletitle{{The Racket Manifesto}}. In
  \bibinfo{booktitle}{\emph{1st Summit on Advances in Programming Languages
  (SNAPL 2015)}} \emph{(\bibinfo{series}{Leibniz International Proceedings in
  Informatics (LIPIcs)}, Vol.~\bibinfo{volume}{32})},
  \bibfield{editor}{\bibinfo{person}{Thomas Ball}, \bibinfo{person}{Rastislav
  Bodik}, \bibinfo{person}{Shriram Krishnamurthi}, \bibinfo{person}{Benjamin~S.
  Lerner}, {and} \bibinfo{person}{Greg Morriset}} (Eds.).
  \bibinfo{publisher}{Schloss Dagstuhl -- Leibniz-Zentrum f{\"u}r Informatik},
  \bibinfo{address}{Dagstuhl, Germany}, \bibinfo{pages}{113--128}.
\newblock
\showISBNx{978-3-939897-80-4}
\showISSN{1868-8969}
\href{https://doi.org/10.4230/LIPIcs.SNAPL.2015.113}{doi:\nolinkurl{10.4230/LIPIcs.SNAPL.2015.113}}


\bibitem[Fickett(1985)]%
        {fickett_introduction_1985}
\bibfield{author}{\bibinfo{person}{Wildon Fickett}.}
  \bibinfo{year}{1985}\natexlab{}.
\newblock \bibinfo{booktitle}{\emph{Introduction to detonation theory}}.
\newblock \bibinfo{publisher}{Univ. of California Pr},
  \bibinfo{address}{Berkeley}.
\newblock
\showISBNx{9780520051256}


\bibitem[Goldberg(1991)]%
        {goldberg_what_1991}
\bibfield{author}{\bibinfo{person}{David Goldberg}.}
  \bibinfo{year}{1991}\natexlab{}.
\newblock \showarticletitle{What every computer scientist should know about
  floating-point arithmetic}.
\newblock \bibinfo{journal}{\emph{Comput. Surveys}} \bibinfo{volume}{23},
  \bibinfo{number}{1} (\bibinfo{date}{March} \bibinfo{year}{1991}),
  \bibinfo{pages}{5--48}.
\newblock
\showISSN{0360-0300, 1557-7341}
\href{https://doi.org/10.1145/103162.103163}{doi:\nolinkurl{10.1145/103162.103163}}


\bibitem[Gorard(2024)]%
        {gorard_gravitas_2024}
\bibfield{author}{\bibinfo{person}{Jonathan Gorard}.}
  \bibinfo{year}{2024}\natexlab{}.
\newblock \bibinfo{title}{Computational General Relativity in the Wolfram
  Language using Gravitas II: ADM Formalism and Numerical Relativity}.
\newblock
\showeprint[arxiv]{2401.14209}~[gr-qc]
\urldef\tempurl%
\url{https://arxiv.org/abs/2401.14209}
\showURL{%
\tempurl}


\bibitem[Gorard et~al\mbox{.}(2024)]%
        {gorard_tetrad_2024}
\bibfield{author}{\bibinfo{person}{Jonathan Gorard}, \bibinfo{person}{Ammar
  Hakim}, \bibinfo{person}{James Juno}, {and} \bibinfo{person}{Jason~M.
  TenBarge}.} \bibinfo{year}{2024}\natexlab{}.
\newblock \bibinfo{title}{A Tetrad-First Approach to Robust Numerical
  Algorithms in General Relativity}.
\newblock
\showeprint[arxiv]{2410.02549}~[gr-qc]
\urldef\tempurl%
\url{https://arxiv.org/abs/2410.02549}
\showURL{%
\tempurl}


\bibitem[Gottlieb et~al\mbox{.}(2001)]%
        {gottlieb_strong_2001}
\bibfield{author}{\bibinfo{person}{Sigal Gottlieb}, \bibinfo{person}{Chi-Wang
  Shu}, {and} \bibinfo{person}{Eitan Tadmor}.} \bibinfo{year}{2001}\natexlab{}.
\newblock \showarticletitle{Strong {Stability}-{Preserving} {High}-{Order}
  {Time} {Discretization} {Methods}}.
\newblock \bibinfo{journal}{\emph{SIAM Rev.}} \bibinfo{volume}{43},
  \bibinfo{number}{1} (\bibinfo{date}{Jan.} \bibinfo{year}{2001}),
  \bibinfo{pages}{89--112}.
\newblock
\showISSN{0036-1445, 1095-7200}
\href{https://doi.org/10.1137/S003614450036757X}{doi:\nolinkurl{10.1137/S003614450036757X}}


\bibitem[Hakim et~al\mbox{.}(2006)]%
        {hakim_high_2006}
\bibfield{author}{\bibinfo{person}{A. Hakim}, \bibinfo{person}{J. Loverich},
  {and} \bibinfo{person}{U. Shumlak}.} \bibinfo{year}{2006}\natexlab{}.
\newblock \showarticletitle{A high resolution wave propagation scheme for ideal
  {Two}-{Fluid} plasma equations}.
\newblock \bibinfo{journal}{\emph{J. Comput. Phys.}} \bibinfo{volume}{219},
  \bibinfo{number}{1} (\bibinfo{date}{Nov.} \bibinfo{year}{2006}),
  \bibinfo{pages}{418--442}.
\newblock
\showISSN{00219991}
\href{https://doi.org/10.1016/j.jcp.2006.03.036}{doi:\nolinkurl{10.1016/j.jcp.2006.03.036}}


\bibitem[Harten(1983)]%
        {harten_high_1983}
\bibfield{author}{\bibinfo{person}{Ami Harten}.}
  \bibinfo{year}{1983}\natexlab{}.
\newblock \showarticletitle{High resolution schemes for hyperbolic conservation
  laws}.
\newblock \bibinfo{journal}{\emph{J. Comput. Phys.}} \bibinfo{volume}{49},
  \bibinfo{number}{3} (\bibinfo{date}{March} \bibinfo{year}{1983}),
  \bibinfo{pages}{357--393}.
\newblock
\showISSN{00219991}
\href{https://doi.org/10.1016/0021-9991(83)90136-5}{doi:\nolinkurl{10.1016/0021-9991(83)90136-5}}


\bibitem[Harten et~al\mbox{.}(1976)]%
        {harten_finitedifference_1976}
\bibfield{author}{\bibinfo{person}{A. Harten}, \bibinfo{person}{J.~M. Hyman},
  \bibinfo{person}{P.~D. Lax}, {and} \bibinfo{person}{B. Keyfitz}.}
  \bibinfo{year}{1976}\natexlab{}.
\newblock \showarticletitle{On finite‐difference approximations and entropy
  conditions for shocks}.
\newblock \bibinfo{journal}{\emph{Communications on Pure and Applied
  Mathematics}} \bibinfo{volume}{29}, \bibinfo{number}{3} (\bibinfo{date}{May}
  \bibinfo{year}{1976}), \bibinfo{pages}{297--322}.
\newblock
\showISSN{0010-3640, 1097-0312}
\href{https://doi.org/10.1002/cpa.3160290305}{doi:\nolinkurl{10.1002/cpa.3160290305}}


\bibitem[Hartman(1960)]%
        {hartman_lemma_1960}
\bibfield{author}{\bibinfo{person}{Philip Hartman}.}
  \bibinfo{year}{1960}\natexlab{}.
\newblock \showarticletitle{A lemma in the theory of structural stability of
  differential equations}.
\newblock \bibinfo{journal}{\emph{Proc. Amer. Math. Soc.}}
  \bibinfo{volume}{11}, \bibinfo{number}{4} (\bibinfo{date}{Aug.}
  \bibinfo{year}{1960}), \bibinfo{pages}{610--620}.
\newblock
\showISSN{0002-9939, 1088-6826}
\href{https://doi.org/10.1090/S0002-9939-1960-0121542-7}{doi:\nolinkurl{10.1090/S0002-9939-1960-0121542-7}}


\bibitem[{IEEE Computer Society} et~al\mbox{.}(2008)]%
        {ieee_computer_society_ieee_2008}
\bibfield{editor}{\bibinfo{person}{{IEEE Computer Society}},
  \bibinfo{person}{{Institute of Electrical and Electronics Engineers}}, {and}
  \bibinfo{person}{{IEEE-SA Standards Board}}} (Eds.).
  \bibinfo{year}{2008}\natexlab{}.
\newblock \bibinfo{booktitle}{\emph{{IEEE} standard for floating-point
  arithmetic}}.
\newblock \bibinfo{publisher}{Institute of Electrical and Electronics
  Engineers}, \bibinfo{address}{New York, NY}.
\newblock
\showISBNx{9780738157528 9780738157535}


\bibitem[Juno(2020)]%
        {juno_2020}
\bibfield{author}{\bibinfo{person}{James Juno}.}
  \bibinfo{year}{2020}\natexlab{}.
\newblock \bibinfo{title}{A Deep Dive into the Distribution Function:
  Understanding Phase Space Dynamics with Continuum Vlasov-Maxwell
  Simulations}.
\newblock
\showeprint[arxiv]{2005.13539}~[physics.plasm-ph]
\urldef\tempurl%
\url{https://arxiv.org/abs/2005.13539}
\showURL{%
\tempurl}


\bibitem[Juno et~al\mbox{.}(2018)]%
        {juno_discontinuous_2018}
\bibfield{author}{\bibinfo{person}{J. Juno}, \bibinfo{person}{A. Hakim},
  \bibinfo{person}{J. TenBarge}, \bibinfo{person}{E. Shi}, {and}
  \bibinfo{person}{W. Dorland}.} \bibinfo{year}{2018}\natexlab{}.
\newblock \showarticletitle{Discontinuous {Galerkin} algorithms for fully
  kinetic plasmas}.
\newblock \bibinfo{journal}{\emph{J. Comput. Phys.}}  \bibinfo{volume}{353}
  (\bibinfo{date}{Jan.} \bibinfo{year}{2018}), \bibinfo{pages}{110--147}.
\newblock
\showISSN{00219991}
\href{https://doi.org/10.1016/j.jcp.2017.10.009}{doi:\nolinkurl{10.1016/j.jcp.2017.10.009}}


\bibitem[Laney(1998)]%
        {laney_computational_1998}
\bibfield{author}{\bibinfo{person}{Culbert~B. Laney}.}
  \bibinfo{year}{1998}\natexlab{}.
\newblock \bibinfo{booktitle}{\emph{Computational {Gasdynamics}}
  (\bibinfo{edition}{1} ed.)}.
\newblock \bibinfo{publisher}{Cambridge University Press}.
\newblock
\showISBNx{9780521570695 9780521625586 9780511605604}
\href{https://doi.org/10.1017/CBO9780511605604}{doi:\nolinkurl{10.1017/CBO9780511605604}}


\bibitem[Lax(1954)]%
        {lax_weak_1954}
\bibfield{author}{\bibinfo{person}{Peter~D. Lax}.}
  \bibinfo{year}{1954}\natexlab{}.
\newblock \showarticletitle{Weak solutions of nonlinear hyperbolic equations
  and their numerical computation}.
\newblock \bibinfo{journal}{\emph{Communications on Pure and Applied
  Mathematics}} \bibinfo{volume}{7}, \bibinfo{number}{1} (\bibinfo{date}{Feb.}
  \bibinfo{year}{1954}), \bibinfo{pages}{159--193}.
\newblock
\showISSN{0010-3640, 1097-0312}
\href{https://doi.org/10.1002/cpa.3160070112}{doi:\nolinkurl{10.1002/cpa.3160070112}}


\bibitem[LeVeque(1992)]%
        {leveque_numerical_1992}
\bibfield{author}{\bibinfo{person}{Randall~J. LeVeque}.}
  \bibinfo{year}{1992}\natexlab{}.
\newblock \bibinfo{booktitle}{\emph{Numerical {Methods} for {Conservation}
  {Laws}}}.
\newblock \bibinfo{publisher}{Birkhäuser Basel}, \bibinfo{address}{Basel}.
\newblock
\showISBNx{9783764327231 9783034886291}
\href{https://doi.org/10.1007/978-3-0348-8629-1}{doi:\nolinkurl{10.1007/978-3-0348-8629-1}}


\bibitem[LeVeque(2011)]%
        {leveque_finite_2011}
\bibfield{author}{\bibinfo{person}{Randall~J. LeVeque}.}
  \bibinfo{year}{2011}\natexlab{}.
\newblock \bibinfo{booktitle}{\emph{Finite volume methods for hyperbolic
  problems} (\bibinfo{edition}{10. printing} ed.)}.
\newblock \bibinfo{publisher}{Cambridge Univ. Press},
  \bibinfo{address}{Cambridge}.
\newblock
\showISBNx{9780521810876 9780521009249}


\bibitem[Munz et~al\mbox{.}(2000)]%
        {munz_divergence_2000}
\bibfield{author}{\bibinfo{person}{C.-D. Munz}, \bibinfo{person}{P. Omnes},
  \bibinfo{person}{R. Schneider}, \bibinfo{person}{E. Sonnendrücker}, {and}
  \bibinfo{person}{U. Voß}.} \bibinfo{year}{2000}\natexlab{}.
\newblock \showarticletitle{Divergence {Correction} {Techniques} for {Maxwell}
  {Solvers} {Based} on a {Hyperbolic} {Model}}.
\newblock \bibinfo{journal}{\emph{J. Comput. Phys.}} \bibinfo{volume}{161},
  \bibinfo{number}{2} (\bibinfo{date}{July} \bibinfo{year}{2000}),
  \bibinfo{pages}{484--511}.
\newblock
\showISSN{00219991}
\href{https://doi.org/10.1006/jcph.2000.6507}{doi:\nolinkurl{10.1006/jcph.2000.6507}}


\bibitem[Musha and Higuchi(1978)]%
        {musha_traffic_1978}
\bibfield{author}{\bibinfo{person}{Toshimitsu Musha} {and}
  \bibinfo{person}{Hideyo Higuchi}.} \bibinfo{year}{1978}\natexlab{}.
\newblock \showarticletitle{Traffic {Current} {Fluctuation} and the {Burgers}
  {Equation}}.
\newblock \bibinfo{journal}{\emph{Japanese Journal of Applied Physics}}
  \bibinfo{volume}{17}, \bibinfo{number}{5} (\bibinfo{date}{May}
  \bibinfo{year}{1978}), \bibinfo{pages}{811--816}.
\newblock
\showISSN{0021-4922, 1347-4065}
\href{https://doi.org/10.1143/JJAP.17.811}{doi:\nolinkurl{10.1143/JJAP.17.811}}


\bibitem[Osher(1984)]%
        {osher_riemann_1984}
\bibfield{author}{\bibinfo{person}{Stanley Osher}.}
  \bibinfo{year}{1984}\natexlab{}.
\newblock \showarticletitle{Riemann {Solvers}, the {Entropy} {Condition}, and
  {Difference} Approximations}.
\newblock \bibinfo{journal}{\emph{SIAM J. Numer. Anal.}} \bibinfo{volume}{21},
  \bibinfo{number}{2} (\bibinfo{date}{April} \bibinfo{year}{1984}),
  \bibinfo{pages}{217--235}.
\newblock
\showISSN{0036-1429, 1095-7170}
\href{https://doi.org/10.1137/0721016}{doi:\nolinkurl{10.1137/0721016}}


\bibitem[Robinson and Voronkov(2001)]%
        {robinson_handbook_2001}
\bibfield{author}{\bibinfo{person}{John~Alan Robinson} {and}
  \bibinfo{person}{Andrei Voronkov}.} \bibinfo{year}{2001}\natexlab{}.
\newblock \bibinfo{booktitle}{\emph{Handbook of automated reasoning}}.
\newblock \bibinfo{publisher}{Elsevier MIT Press}, \bibinfo{address}{Amsterdam
  New York Cambridge, Mass}.
\newblock
\showISBNx{9780444508133}


\bibitem[Roe(1981)]%
        {roe_approximate_1981}
\bibfield{author}{\bibinfo{person}{P.L Roe}.} \bibinfo{year}{1981}\natexlab{}.
\newblock \showarticletitle{Approximate {Riemann} solvers, parameter vectors,
  and difference schemes}.
\newblock \bibinfo{journal}{\emph{J. Comput. Phys.}} \bibinfo{volume}{43},
  \bibinfo{number}{2} (\bibinfo{date}{Oct.} \bibinfo{year}{1981}),
  \bibinfo{pages}{357--372}.
\newblock
\showISSN{00219991}
\href{https://doi.org/10.1016/0021-9991(81)90128-5}{doi:\nolinkurl{10.1016/0021-9991(81)90128-5}}


\bibitem[Roe(1986)]%
        {roe_characteristic-based_1986}
\bibfield{author}{\bibinfo{person}{P~L Roe}.} \bibinfo{year}{1986}\natexlab{}.
\newblock \showarticletitle{Characteristic-{Based} {Schemes} for the {Euler}
  {Equations}}.
\newblock \bibinfo{journal}{\emph{Annual Review of Fluid Mechanics}}
  \bibinfo{volume}{18}, \bibinfo{number}{1} (\bibinfo{date}{Jan.}
  \bibinfo{year}{1986}), \bibinfo{pages}{337--365}.
\newblock
\showISSN{0066-4189, 1545-4479}
\href{https://doi.org/10.1146/annurev.fl.18.010186.002005}{doi:\nolinkurl{10.1146/annurev.fl.18.010186.002005}}


\bibitem[Sweby(1984)]%
        {sweby_high_1984}
\bibfield{author}{\bibinfo{person}{P.~K. Sweby}.}
  \bibinfo{year}{1984}\natexlab{}.
\newblock \showarticletitle{High {Resolution} {Schemes} {Using} {Flux}
  {Limiters} for {Hyperbolic} {Conservation} {Laws}}.
\newblock \bibinfo{journal}{\emph{SIAM J. Numer. Anal.}} \bibinfo{volume}{21},
  \bibinfo{number}{5} (\bibinfo{date}{Oct.} \bibinfo{year}{1984}),
  \bibinfo{pages}{995--1011}.
\newblock
\showISSN{0036-1429, 1095-7170}
\href{https://doi.org/10.1137/0721062}{doi:\nolinkurl{10.1137/0721062}}


\bibitem[Toro(2009)]%
        {toro_riemann_2009}
\bibfield{author}{\bibinfo{person}{E.~F. Toro}.}
  \bibinfo{year}{2009}\natexlab{}.
\newblock \bibinfo{booktitle}{\emph{Riemann solvers and numerical methods for
  fluid dynamics: a practical introduction} (\bibinfo{edition}{3rd ed} ed.)}.
\newblock \bibinfo{publisher}{Springer}, \bibinfo{address}{Dordrecht New York}.
\newblock
\showISBNx{9783540498346}


\bibitem[Toro and Billett(2000)]%
        {toro_centred_2000}
\bibfield{author}{\bibinfo{person}{E.~F. Toro} {and} \bibinfo{person}{S.~J.
  Billett}.} \bibinfo{year}{2000}\natexlab{}.
\newblock \showarticletitle{Centred {TVD} schemes for hyperbolic conservation
  laws}.
\newblock \bibinfo{journal}{\emph{IMA J. Numer. Anal.}} \bibinfo{volume}{20},
  \bibinfo{number}{1} (\bibinfo{date}{Jan.} \bibinfo{year}{2000}),
  \bibinfo{pages}{47--79}.
\newblock
\showISSN{0272-4979, 1464-3642}
\href{https://doi.org/10.1093/imanum/20.1.47}{doi:\nolinkurl{10.1093/imanum/20.1.47}}


\bibitem[Van~Leer(1974)]%
        {van_leer_towards_1974}
\bibfield{author}{\bibinfo{person}{Bram Van~Leer}.}
  \bibinfo{year}{1974}\natexlab{}.
\newblock \showarticletitle{Towards the ultimate conservative difference
  scheme. {II}. {Monotonicity} and conservation combined in a second-order
  scheme}.
\newblock \bibinfo{journal}{\emph{J. Comput. Phys.}} \bibinfo{volume}{14},
  \bibinfo{number}{4} (\bibinfo{date}{March} \bibinfo{year}{1974}),
  \bibinfo{pages}{361--370}.
\newblock
\showISSN{00219991}
\href{https://doi.org/10.1016/0021-9991(74)90019-9}{doi:\nolinkurl{10.1016/0021-9991(74)90019-9}}


\bibitem[Van~Leer(1977)]%
        {van_leer_towards_1977}
\bibfield{author}{\bibinfo{person}{Bram Van~Leer}.}
  \bibinfo{year}{1977}\natexlab{}.
\newblock \showarticletitle{Towards the ultimate conservative difference scheme
  {III}. {Upstream}-centered finite-difference schemes for ideal compressible
  flow}.
\newblock \bibinfo{journal}{\emph{J. Comput. Phys.}} \bibinfo{volume}{23},
  \bibinfo{number}{3} (\bibinfo{date}{March} \bibinfo{year}{1977}),
  \bibinfo{pages}{263--275}.
\newblock
\showISSN{00219991}
\href{https://doi.org/10.1016/0021-9991(77)90094-8}{doi:\nolinkurl{10.1016/0021-9991(77)90094-8}}


\bibitem[Van~Leer(1979)]%
        {van_leer_towards_1979}
\bibfield{author}{\bibinfo{person}{Bram Van~Leer}.}
  \bibinfo{year}{1979}\natexlab{}.
\newblock \showarticletitle{Towards the ultimate conservative difference
  scheme. {V}. {A} second-order sequel to {Godunov}'s method}.
\newblock \bibinfo{journal}{\emph{J. Comput. Phys.}} \bibinfo{volume}{32},
  \bibinfo{number}{1} (\bibinfo{date}{July} \bibinfo{year}{1979}),
  \bibinfo{pages}{101--136}.
\newblock
\showISSN{00219991}
\href{https://doi.org/10.1016/0021-9991(79)90145-1}{doi:\nolinkurl{10.1016/0021-9991(79)90145-1}}


\bibitem[Vreugdenhil(1994)]%
        {vreugdenhil_numerical_1994}
\bibfield{author}{\bibinfo{person}{C.~B. Vreugdenhil}.}
  \bibinfo{year}{1994}\natexlab{}.
\newblock \bibinfo{booktitle}{\emph{Numerical {Methods} for {Shallow}-{Water}
  {Flow}}}. \bibinfo{series}{Water {Science} and {Technology} {Library}},
  Vol.~\bibinfo{volume}{13}.
\newblock \bibinfo{publisher}{Springer Netherlands},
  \bibinfo{address}{Dordrecht}.
\newblock
\showISBNx{9789048144723 9789401583541}
\href{https://doi.org/10.1007/978-94-015-8354-1}{doi:\nolinkurl{10.1007/978-94-015-8354-1}}


\bibitem[Wang et~al\mbox{.}(2020)]%
        {wang_exact_2020}
\bibfield{author}{\bibinfo{person}{Liang Wang}, \bibinfo{person}{Ammar~H.
  Hakim}, \bibinfo{person}{Jonathan Ng}, \bibinfo{person}{Chuanfei Dong}, {and}
  \bibinfo{person}{Kai Germaschewski}.} \bibinfo{year}{2020}\natexlab{}.
\newblock \showarticletitle{Exact and locally implicit source term solvers for
  multifluid-{Maxwell} systems}.
\newblock \bibinfo{journal}{\emph{J. Comput. Phys.}}  \bibinfo{volume}{415}
  (\bibinfo{date}{Aug.} \bibinfo{year}{2020}), \bibinfo{pages}{109510}.
\newblock
\showISSN{00219991}
\href{https://doi.org/10.1016/j.jcp.2020.109510}{doi:\nolinkurl{10.1016/j.jcp.2020.109510}}


\bibitem[Wesseling(2001)]%
        {wesseling_principles_2001}
\bibfield{author}{\bibinfo{person}{Pieter Wesseling}.}
  \bibinfo{year}{2001}\natexlab{}.
\newblock \bibinfo{booktitle}{\emph{Principles of {Computational} {Fluid}
  {Dynamics}}}. \bibinfo{series}{Springer {Series} in {Computational}
  {Mathematics}}, Vol.~\bibinfo{volume}{29}.
\newblock \bibinfo{publisher}{Springer Berlin Heidelberg},
  \bibinfo{address}{Berlin, Heidelberg}.
\newblock
\showISBNx{9783642051456 9783642051463}
\href{https://doi.org/10.1007/978-3-642-05146-3}{doi:\nolinkurl{10.1007/978-3-642-05146-3}}


\end{thebibliography}
